\documentclass[prc,twocolumn,showpacs,preprintnumbers,amsmath,amssymb]
{revtex4}

\usepackage{graphicx}
\usepackage{dcolumn}
\usepackage{bm}

\begin{document}


\title{Decoding Beta--Decay Systematics:\\
A Global Statistical Model for $\beta^-$ Halflives}\homepage{http://www.pythaim.phys.uoa.gr}

\author{N.~J.~Costiris}
\email[ E-mail: ]{ncost@phys.uoa.gr, nick.costiris@gmail.com}
\author{E. Mavrommatis}
\email[ E-mail: ]{emavrom@phys.uoa.gr}
\affiliation{Physics Department, Section of Nuclear \&
Particle Physics\\
University of Athens, GR-15771 Athens, Greece}
\author{K. A. Gernoth}
\email[ E-mail: ]{klaus.a.gernoth@@manchester.ac.uk}
\affiliation{
Institut f\"ur Theoretische Physik, Johannes-Kepler-Universit\"at,
A-4040 Linz, Austria}
\affiliation{School of Physics \& Astronomy, Schuster Building\\ The
University of Manchester, Manchester, M13 9PL, UK}
\author{J.~W.~Clark}
\email[ E-mail: ]{jwc@wustl.edu}
\affiliation{McDonnell Center for the Space Sciences \&
Department of Physics\\ Washington University,
St.~Louis, MO 63130, USA,}
\affiliation{Complexo Interdisciplinar, Centro de
Mathem\'atica e Aplica\c{c}\~{o}es Fundamentals\\
University of Lisbon, 1649-003 Lisbon, Portugal}
\affiliation{Departamento de F\'isica, Instituto Superior T\'ecnico,\\
Technical University of Lisbon, 1096 Lisbon, Portugal\\}

\date{May 2008, Submitted to Phys. Rev. C}

\begin{abstract}

Statistical modeling of nuclear data provides a novel approach to nuclear systematics  complementary to established theoretical and phenomenological approaches based on quantum theory.  Continuing previous studies in which global statistical modeling is pursued within the general framework of machine learning theory, we implement advances in training algorithms designed to improved generalization, in application to the problem of reproducing and predicting the halflives of nuclear ground states that decay 100\% by the $\beta^-$ mode.  More specifically,
fully-connected, multilayer feedforward artificial neural network models are developed  using the Levenberg-Marquardt optimization algorithm together with Bayesian regularization and cross-validation.  The predictive performance of models emerging from extensive computer experiments is compared with that of traditional microscopic and phenomenological
models as well as with the performance of other learning systems, including earlier neural network models as well as the support vector machines recently applied to the same problem. 
In discussing the results,  emphasis is placed on predictions for nuclei that are far from the stability line, and especially those involved in the r-process nucleosynthesis. It is found that the new statistical models can match or even surpass the predictive performance of conventional models for beta-decay systematics and accordingly should provide a valuable additional tool for exploring the expanding nuclear landscape.
\end{abstract}


\maketitle

\section{\label{sec:level1}INTRODUCTION}
\begin{quotation} 
\small \it
``Numbers are the within of all things.''

Pythagoras of Samos
\end{quotation} 

This work is devoted to the development of artificial neural
network models which, after being trained with a subset of
the available experimental data on beta decay from nuclear
ground states, demonstrate significant reliability in the
prediction of $\beta^-$ halflives for nuclides absent
from the training set.  The work represents an exploratory
study of the degree to which the existing data determines
the mapping from proton and neutron numbers to the
corresponding $\beta^-$ halflife.

There is an urgent need among nuclear physicists and astrophysicists for reliable estimates of  $\beta^-$-decay halflives of nuclei far from stability~\cite{A1,eirhnh2}.   Among nuclear physicists this need is driven both by the experimental programs of existing and future radioactive ion beam facilities and by the stresses placed on established nuclear structure theory  as totally new areas of the nuclear landscape are opened for exploration. For nuclear astrophysicists, such information is intrinsic to an understanding of 
supernova explosions -- the initialization of the explosion, the subsequent neutronization of the core material, and the strength and fate of the shock wave formed -- and the nucleosynthesis of heavy elements above \rm{Fe}, notably the r-process~\cite{eirhnh3,eirhnh4,eirhnh5}. Both the element distribution on the r-path and the time scale of the r-process are highly sensitive to the $\beta$-decay properties of the neutron-rich nuclei involved.   

In the nuclear chart there are spaces for some 6000 nuclides between the
$\beta$-stability line and the neutron-drip line. Except for a few key
nuclei, $\beta$ decay of r-process nuclei cannot be studied in terrestrial
laboratories, so the required information must come from nuclear models.
Over the years, a number of approaches for modeling of $\beta^-$-decay
halflives have been proposed and applied.  These include the more
phenomenological treatments, such as the Gross Theory (GT), as well as
microscopic approaches based on the shell model and the proton-neutron
Quasiparticle Random-Phase Approximation ($pn$QRPA) in various versions.
More recently, hybrid macroscopic-microscopic and
relativistic models have come on the scene.  Some of
these approaches emphasize only global applicability, while
others seek self-consistency or comprehensive
inclusion of nuclear correlations.  Table 1 of Ref.~\onlinecite{111}
provides a convenient summary of a number of the competing models
of beta-decay systematics.

In Gross Theory, developed by Takahashi, Yamada and Kondoh~\cite{22}, gross properties of $\beta^-$ decay over a wide nuclidic region are predicted by averaging over the final states of the daughter nucleus.  Subsequently, various refinements and modifications of this treatment have been introduced.   The most current of these is the so-called Semi-Gross Theory (SGT), in which the shell effects of only the parent nucleus are taken into account~\cite{12}.  On the other hand, in the calculations of $\beta^-$-decay halflives within the shell model, the detailed structure of  $\beta$ strength function is considered.  Results exist for lighter nuclei and nuclei at  $N= 50, 82$, and $126$.  (See Refs.~\onlinecite{eirhnh9, eirhnh10}
for recent calculations.)  Due to the limits set by the size of the configuration space, calculations are not possible for heavy nuclei.

Several groups have carried out extensive $pn$QRPA studies including pairing.  Efforts along this line
by Klapdor and co-workers~\cite{eirhnh11} began in the framework of the Nilsson single-particle model, including the Gamow-Teller residual interaction in Tamm-Dancoff approximation (TDA), with pairing treated at the BCS level \cite{2}.  This approach has been complemented and refined by 
Staudt et ~al.~\cite{5} and Hirsch et al.~\cite{eirhnh14}, using $pn$QRPA with the Gamow-Teller residual interaction. The later study by Homma et al.~\cite{6}, denoted NBCS + $pn$QRPA, includes a 
schematic interaction also for the first-forbidden (\textit{ff}) decay.  The Klapdor group has extended the $pn$QRPA theory to calculate $\beta$-decay halflives in stellar environments using configurations 
beyond 1$p-$1$h$ \cite{eirhnh16}.

The starting point of the $\beta$-decay calculations of M\"{o}ller and co-workers is the study
of nuclear-ground-state masses and deformations based on the finite-range droplet model (FRDM) and a folded-Yukawa single-particle potential~\cite{24}. The  $\beta$-decay halflives for the allowed Gamow-Teller transitions have been obtained from a $pn$QRPA calculation after the addition of pairing and Gamow-Teller residual interactions, in a procedure denoted FRDM + $pn$QRPA~\cite{eirhnh18, 7}. In the latest calculations the effect of the \textit{ff} decay has been added by using the Gross Theory 
($pn$QRPA +\textit{ff}GT)~\cite{8}.  Non-relativistic $pn$QRPA calculations that aim at self-consistency include the Hartree-Fock-Bogoliubov + continuum QRPA (HFB + QRPA) calculations performed with a Skyrme energy-density functional for some spherical even-even semi-magic nuclides with   $N= 50, 82, 126$~\cite{eirhnh21}. The extended Thomas-Fermi plus Strutinski integral method (ETFSI) (an approximation to HF method based on a  
Skyrme-type force plus a  $\delta-$function pairing force) has been elaborated and applied to large-scale predictions of $\beta^-$  halflives~\cite{56}. Recently, the density functional + continuum QRPA (DF + CQRPA) approximation, with the spin-isospin effective \rm{NN} interaction of the finite Fermi system theory operating in the \textit{ph} channel, has been developed for ground-state properties and Gamow-Teller and \textit{ff} transitions of nuclei far from the stability line, and applied near closed neutron shells at $N= 50, 82, 126$  and in the region ``east'' of $^{208}{\rm Pb}$~\cite{eirhnh23,111}. In the relativistic framework, a $pn$QRPA calculation 
($pn$RQPRA) based on a relativistic 
Hartree-Bogoliubov description of nuclear ground states with the density-dependent effective interaction
DD-MEI* has been employed to obtain Gamow-Teller $\beta^-$-decay halflives of neutron-rich nuclei in the $N\simeq 50$   and $N\simeq 82$ regions relevant to the r-process~\cite{57}. Recently, an extension of the above framework to include momentum-dependent nucleon self-energies was applied in the calculation of  $\beta$-decay halflives of neutron-rich nuclei in the  $Z\simeq 28$ and  $Z\simeq 50$ regions~\cite{eirhnh77}.

Despite continuing methodological improvements,  the predictive power of these conventional, ``theory-thick'' models is rather limited for  $\beta^-$-decay halflives of nuclei that are mainly far from stability.  The
predictions often deviate from experiment by one or more orders of magnitude and show considerable sensitivity to quantities that are poorly known. In this environment, statistical modeling based on advanced 
techniques of statistical learning theory or ``machine-learning,'' notably artificial  neural networks (ANNs)~\cite{19,15} and support vector machines 
(SVMs)~\cite{15,27,eirhnh28}, offers an interesting and potentially effective alternative for global modeling of  $\beta^-$-decay lifetimes.  Such approaches have proven their value for a variety of scientific problems in astronomy, high-energy physics, and biochemistry that involve function approximation and pattern classification~\cite{59,60}. 
Statistical modeling implementing machine-learning algorithms is ``theory-thin,'' since it is driven by data with minimal guidance from mechanistic concepts; thus it is very different from the ``theory-thick'' approaches summarized above. Any nuclear observable $X$ can be viewed as a mapping from the atomic and neutron numbers $Z$ and $N$ identifying an arbitrary nuclide, to the corresponding value of the observable (the $\beta$ halflife, in the present study).  In machine learning, one attempts to approximate the mapping $(Z,N)\to X$ based only on an available subset of the data for X, i.e., a body of {\it training data} consisting of known examples of the mapping.  One attempts to {\it infer} the mapping, in the sense of Bayesian probability theory as expounded by Jaynes~\cite{eirhnh31}. Thus, one is asking the question: ``To what extent does the data, and only the data, determine the mapping $(Z,N)\to X$?''  The answer (or answers) to this question should surely be of fundamental interest, when confronted with databases as large, complex, and refined as those existing in nuclear physics.  

A learning machine consists of (i) an input interface where, for example, input variables $Z$ and $N$ are fed to the device in coded form, (ii) a system of intermediate elements or units that process the input, and (iii) an output interface where an estimate of the corresponding observable of interest, say the beta halflife $T_{\beta}$ appears for decoding.
Given an adequate body of training data (consisting of input ``patterns'' or vectors and their appropriate outputs), a suitable learning algorithm is used to adjust the parameters of the machine, e.g., the weights of the connections between the processing elements in the case of a neural network. These parameters are adjusted in such a way that the learning machine (a) generates responses at the output interface that closely fit the halflives of the training examples and (b) serves as a reliable predictor of the halflives of the test nuclei absent from the training set. In the more mundane language of function approximation, the learning-machine model provides a means for \textit{interpolation} or \textit{extrapolation}. 

Neural-network models have already been constructed for a range of nuclear properties including atomic masses, neutron separation energies, ground state spins and parities, and branching probabilities  for different decay channels, as well as  $\beta^-$-decay halflives~\cite{59,60,13,14,70,29}. Very recently, global statistical models of some of these properties have also been developed based on support vector machines~\cite{25,55,eirhnh38}.  In time, there has been steady improvement of the quality of these models, such that the documented performance of the best examples approaches or even surpasses that of the traditional ``theory-thick'' models in predictive reliability.  By their nature, they should not be expected to compete with traditional phenomenological or microscopic models in generating new physical insights. However, their prospects for revealing new regularities are by no means sterile, since the explicit formula created by the learning algorithm for the physical observable being modeled is available for analysis.

We present here a new global model for the halflives of nuclear ground states that decay 100\% by the $\beta^-$ mode, developed by implementing the most recent advances in machine-learning algorithms.   Sec.~II describes the elements of the model, the training algorithm employed, steps taken to improve generalization, the data sets adopted, and the coding schemes used at input and output interfaces. Performance measures for assessing the quality of global models of beta lifetimes are reviewed in Sec. III. The results of our large-scale modeling studies are reported and evaluated in Sect. IV. Detailed comparisons are made with experiment, with a selection of the theory-driven GT and $pn$QRPA global models, and with previous ANN and SVM models. This assessment is followed by the presentation of specific predictions for nuclei that are situated far from the line of stability, focusing in particular at those involved in r-process nucleosynthesis. Finally, Sect. V summarizes the conclusions of the present study and considers
the prospects for further improvements in statistical prediction of halflives.

\section{\label{sec:level2}THE MODEL}

\subsection{\label{sec:leve21}Network Architecture and Dynamics}

\begin{figure}
\includegraphics{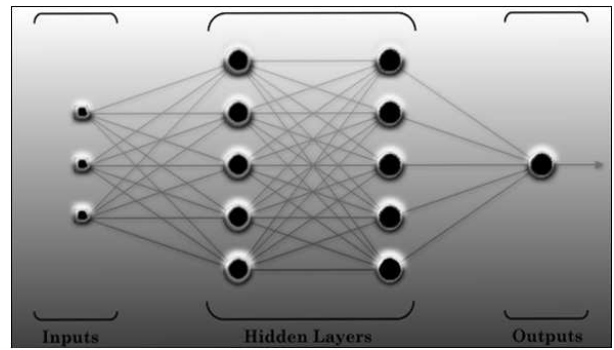}
\caption{\label{fig:one} Architecture of  a typical fully connected feedforward network having an input layer with three units, two hidden layers each containing five units, and a single output unit, thus of structure $\left[ {3-5-5-1 \left| 56  \right.} \right]$.}
\end{figure}

\textit{Artificial neural networks}, whose structure is inspired by the anatomy of natural neural systems,  consist of interconnected dynamical units (sometimes called neurons) that are typically arranged in a distinct layered topology.  Also in analogy with biological neural systems, the function of the network, for example pattern recognition, is determined by the connections between the units. In the work to be reported, we have focused exclusively on feedforward networks, in which information flows unidirectionally from an input layer through one or more intermediate (hidden) layers to an output layer.  Lateral and feedback connections are absent, but otherwise the network is fully connected.   The activation of hidden units is nonlinear, whereas the output units transform their inputs linearly. The architecture of such a network is indicated by the notation
\begin{equation}
\left[ {I - H_1  - H_2  -  \cdots  - H_L  - O \left| {W} \right.} \right],
\label{eq:1}
\end{equation}
where $I$ is the number of inputs, $H_i$ is the number of neurons in the $i^{th}$ hidden layer, $O$ is the number of units in the output layer, and $W$ is the total number of parameters needed to complete the specification of the network, consisting of the weights of the connections and the biases of the units.  Fig.~\ref{fig:one} depicts a typical fully connected network of the class used in our  statistical modeling, in this case having architecture $\left[ {3-5-5-1 \left| 56  \right.} \right]$.

The connection from neuron $j$ to neuron $i$ carries a real-number weight $w_{ij}$.   Thus, if $o_j$ is the activity of neuron $j$, it provides an input $w_{ij} o_j$ to neuron $i$. In addition, each neuron $i$ is assigned a bias parameter $b_i$, which is summed together with its input signals from other neurons $j$ to form its total input $u_i$. This quantity is fed into the activation function $\varphi_i$ characterizing the response of neuron $i$.  For the neurons in hidden layers, this function is 
taken to have the nonlinear \textit{hyperbolic tangent} form 
\begin{equation}
\varphi \left( u \right) = \frac{2}{{1 + \exp \left( { - 2u} \right)}} - 1,
\label{eq:tanh}
\end{equation}
while for the neurons in the output layer  the \textit{symmetric saturating linear} form 
\begin{eqnarray}
\varphi \left( u \right) = \left\{ {\begin{array}{*{20}r}
   {- 1,}  \\
   u,  \\
   1,  \\
\end{array}\begin{array}{*{20}c}
   {u <  - 1}  \\
   { - 1 \le u \le 1},  \\
   {u > 1}  \\
\end{array}} \right.
\label{eq:satlin}
\end{eqnarray}
is adopted. The {\it output} (or {\it activity}) $o_i$ of neuron $i$ is given by
\begin{equation}
o_i = \varphi\left( b_i + \sum\nolimits_j {w_{ij} o_j } \right).
\label{output}
\end{equation}
We note that with its sign reversed, a neuron's bias can be viewed as a threshold for its activation. Also, it is sometimes convenient to regard the bias  $b_i$ as the {\it weight} of a connection to neuron $i$ from a virtual unit $v$ that is always fully ``on'', i.e., $o_v \equiv 1$.  The weights $w_{ij}$ and 
biases $b_i$ are \textit{adjustable} scalar parameters of the untrained network, available for optimization of the network's performance in some task, notably classification and function approximation in the case of applications to global nuclear modeling.  This is usually done by minimizing some measure of the errors made by the network in response to inputs corresponding to a set of training examples, or ``training patterns.''

The dynamics of the network is exceptionally simple. When a pattern $p$ is presented at the 
input, the system computes a response according to two rules:

(a) The states of all neurons within a given layer, as specified by the outputs $o_i$ of
Eq.~(\ref{output}), are updated in parallel, and

(b) The layers are updated successively, proceeding from the input to the output layer.

In modeling the systematics of beta lifetimes with this approach, we apply a supervised learning algorithm to optimize the weights and biases, as described in the subsections to follow. The patterns $p$ to be learned or predicted, examples of the mapping from nuclide to lifetime, take the form
 \begin{eqnarray}
\begin{array}{l}
 \left\{ {\left( {\begin{array}{*{20}c}
   {Z_p}  \\
   {N_p}  \\
  \end{array}} \right),\, \log _{10} {T_{\beta,{\rm exp} }^{p}} } \right\}, \\ 
 \end{array}
\label{eq:5}
\end{eqnarray}
and thus consist of an association between the atomic and neutron numbers
of the parent nuclide, with the base-10 log of the experimental halflife $T_{\beta,{\rm exp}}^{p}$. It is of course natural to work with the logarithm of $T_\beta$, since the observed values of $T_\beta$ itself vary over many orders of magnitude. 

According to the nature of statistical estimation, realized here in the 
application of machine learning techniques to function approximation,
a neural network model is only one form in which empirical knowledge of a physical phenomenon of interest ($\beta$ decay in this case) may be 
encoded~\cite{15}.  As indicated in the introduction, the present work is at some level an investigation of the degree to which the available data determines the physical mapping from $Z$ and $N$ to the corresponding 
$\beta$-decay halflife.  Actually, we do not have knowledge of the exact
functional relationship involved. 
Thus we should write 
\begin{equation}
\log_{10} T_\beta (Z,N) = g(Z,N) + \varepsilon(Z,N),
\label{eq:6}
\end{equation}
where $g(Z,N)$ is a function that decodes the decay systematics and  
$\varepsilon$ is a random expectation error -- a Gaussian noise term that represents our ignorance about the dependence of $T_{\beta }$ on 
$Z$ and $N$.
From a heuristic perspective beyond strict mathematical definitions,
this ${\varepsilon}$ noise term could reflect ``chaotic'' influences on the 
phenomenon, along with missing regularities that could be more easily
modeled and eventually included in the estimate of the physical
quantity $T_\beta$.

The pragmatic objective of the training process in this application will be to minimize the sum of squared errors $e_p$ committed by the network model relative to experiment, for the $n$ patterns $p$ from the available experimental data (D) that constitute the training set
\begin{equation}
{E}_{D}  = \sum\limits_{{{p = 1}}}^{{n}} {\left(e_p \right)}^2  = \sum\limits_{{{p = 1}}}^{{n}} {\left( {{{\log}}_{{{10}}} {{{T}}_{{{\beta }}{{,{\rm exp}}}}^{{p}} }{{ - \log}}_{{{10}}} {{{T}}_{{{\beta }}{{,{\rm calc}}}}^{{p}} } } \right)} ^2 .
\label{eq:7}
\end{equation}
Here ${{\log}}_{{{10}}} {{{T}}_{{{\beta }}{{,{\rm calc}}}}^{{p}} }$ is the neural-network output for pattern (nuclide) $p$, whereas ${\log}_{10}{{{T}}_{{{\beta }}{{,{\rm exp}}}}^{{p}} }$ is the target output.
This quantity is often referred to as a \textit{cost function} or {\it objective function} and can obviously be used as a measure of network performance.  In practice, its form will be modified in Subsec.~C.2 below so as   
to improve the network's ability to generalize, or predict.   A network model is said to generalize well if it performs well for inputs (nuclides) outside the training set, with the mean-square error for these ``fresh'' nuclei providing an appropriate measure of predictive performance.

 \subsection{\label{sec:level23}The Training Algorithm}

In supervised learning, the network is exposed, in succession, to the input patterns (nuclides) of the training set, and the errors made by the network are recorded.  One pass through the training set is called an {\it epoch}.  In {\it batch} training, weights
and biases are incremented after each epoch according to a suitable learning algorithm, with the expectation of improving subsequent performance on the training set.

Statistical modeling inevitably involves a tradeoff between closely fitting the training data and reliability in interpolation and extrapolation \cite{15,27}. 
In the present application, it is not the goal of network training to achieve an exact reproduction, by the model, of
the known halflives.  This would necessarily entail fitting the data precisely with a large number of parameters -- which would in 
general require a complex ANN with many layers and/or 
neurons/layer.  Obviously, there is no point in constructing a lookup table of the known beta halflives.  Rather, the goal is to achieve an
accurate representation of the regularities inherent in the training
data by means of a network that is no more complicated than it need be, thereby promoting good generalization.   

We employ a training algorithm within the general class of backpropagation learning 
procedures.  There are now quite a number of well-tested procedures in this class, 
including steepest-descent, conjugate-gradient, Newton, and  
Levenberg-Marquardt training algorithms~\cite{19}.  All of these 
approaches aim to minimize an appropriate cost function with respect to the network weights and biases.  The term backpropagation refers to the process by which derivatives of network errors with respect to weights/biases can be computed starting from the output layer and proceeding backwards toward the input. In general, the Levenberg-Marquardt backpropagation (LMBP) algorithm will have the fastest convergence in function approximation problems, an advantage that is especially noticeable if very accurate training is required~\cite{1779}.
   
In the Newton method, minimization of the cost function is accomplished through the update rule
\begin{equation}
{\bf {w}}_{k+1}  = {\bf {w}}_k - {\bf {H}}^{-1}_k {\bf {g}}_k,
\label{eq:A10}
\end{equation}
where ${\bf w}_k$ is the vector formed from the weights and biases, ${\bf {H}}_k$ is the Hessian matrix (the matrix of second derivatives of the objective function $E_D$ with respect to  the weights and biases) and ${\bf {g}}_k$ is the gradient of $E_D$ at the current epoch $k$.  
As a Newton-based procedure attempting to approximate the Hessian matrix, the Levenberg-Marquardt algorithm~\cite{18,19} was designed to approach second-order training speed without having to compute second derivatives.  When the cost function has the form of Eq.~(\ref{eq:7}), the Hessian matrix for nonlinear networks can be approximated as 
\begin{equation}
{\bf {H}} \approx {\bf {J}}^{\rm T} {\bf {J}},
\label{eq:8}
\end{equation}
where ${J}$ is the Jacobian matrix
composed of the first derivatives of the network errors with respect to the weights/biases. This generates a $W \times W$ matrix, where $W$ is the number of the free parameters (weights and biases) of the network. The gradient $\bf{g}$ can be computed as 
\begin{equation}
{\bf {g}} = {\bf {J}}^{\rm T} {\bf {e}},
\label{eq:9}
\end{equation}
where ${\bf e}$ is the vector whose components are the network errors $e_p$. 
(As in Eq.~(\ref{eq:7}), the network error for a given input pattern is
the target value of the estimated quantity, minus the value
produced by the network.)

Adopting the Gauss-Newton approximation
(\ref{eq:8}), the Levenberg-Marquardt algorithm then adjusts the weights according to the Newton-like updating rule \begin{equation}
{\bf {w}}_{k+1}  = {\bf {w}}_k - \left[ {{\bf {J}}^{\rm T}_k {\bf {J}}_k + \mu_k {\bf{I}}} \right]^{ - 1} {\bf {J}}^{\rm T}_k {\bf {e}}_k,
\label{eq:10}
\end{equation}
where $\bf{I}$ is the unit matrix.

The factor $\mu_k$ appearing in the Eq.~(\ref{eq:10}) is an adjustable parameter that controls the step size so as to quench oscillations of the cost function near its minimum.  When $\mu_k$ is very small, LMBP coincides with the Newton method executed with the approximate Hessian matrix. When $\mu_k$ is large enough, matrix $\bf{g}$ in Eq.~(\ref{eq:9}) is nearly diagonal and the algorithm behaves like a steepest-descent method with a small step size. Steepest-descent algorithms are based on linear approximation of the cost function, while the Newton algorithm involves quadratic approximation.  Newton's method is faster and more accurate near an error minimum. Therefore the preferred strategy is to shift toward Newton's method as quickly as possible.  To this end, $\mu_k$ is decreased after each successful step and is increased only when a tentative step would raise the cost function. In this way, the cost function will always be reduced at each iteration of the algorithm. 
The algorithm begins with $\mu_k$  set  to some small value
(e.g.,  $\mu_k=0.01$).  If a step does not yield a smaller value for the cost function, the step is repeated with $\mu_k$ multiplied by some factor $\theta  > 1$ (e.g., $\theta  = 10$). Eventually the cost function should decrease. If a step does produce a smaller value for the cost function, then $\mu_k$ is divided by  $\theta$ for the next step, so that the algorithm will approach Gauss-Newton, which should provide faster convergence.   Thus, the Levenberg-Marquardt algorithm is advantageous in implementing a favorable compromise between slow but  guaranteed convergence far from the
minimum and a fast convergence in the neighborhood of the minimum.

The key step in LMBP algorithm is the computation of the Jacobian matrix. To perform this computation we use a variation of the classical backpropagation algorithm.   In the standard backpropagation procedure, one computes the derivatives of the squared errors with respect to the weights and biases of the network. To create the Jacobian matrix we need to compute the derivatives of the errors, instead of the derivatives of their squares, a trivial difference computationally.

\subsection{\label{sec:level24} Improving Generalization}

To build a viable statistical model, it is imperative to avoid the phenomenon of {\it overfitting}, which for example occurs when, under excessive training, the network simply ``memorizes'' the training data and makes a lookup table.  Such a network fails to learn the regularities of the target mapping that are inherent in the data; the network is therefore deficient in generalization. 
We seek to avoid overfitting through a combination of well-established techniques, namely {\it cross-validation} \cite{15} and {\it Bayesian regularization} \cite{17}.

\subsubsection{\label{subsubsec:level241}Cross-Validation}

Cross-validation is a standard statistical technique based on dividing the data into three 
subsets~\cite{15}. The first subset is the \textit{learning} or \textit{training set} employed in building the model (i.e., in computing the Jacobians and updating the network weights and biases). The second subset is the \textit{validation set}, used to evaluate the performance of the model outside the training set and guide the choice of model. The error on the validation set is monitored during the training process. When the network begins to overfit the data, the error on the validation set will typically begin to rise. If this continues to occur for a specified number of iterations, the training is stopped, and the weights and biases at the minimum of the validation error are reinstated.  The third subset is the \textit{test set}. The error on the test set is not used during the training procedure, but it is used to assess the generalization performance of the model and to compare different models.  While effective in suppressing overfitting, cross-validation tends to produce networks whose response is not sufficiently smooth.
This is dealt with by performing Bayesian regularization together with cross-validation.

\subsubsection{\label{subsubsec:level25}Bayesian regularization}

The standard Levenberg-Marquardt algorithm aims to reduce the sum of squared errors $E_D$, 
written explicitly in Eq.~(\ref{eq:7}) for the $\beta$-decay problem. However, in the framework of Bayesian regularization~\cite{17}, the Levenberg-Marquardt {\it optimization} (backpropagation) 
algorithm (denoted LMOBP) minimizes a linear combination of squared errors and squared network parameters,
\begin{equation}
{{F = \tilde{\beta} E}}_{{D}} {{ + \tilde{\alpha} E}}_{{W}}, 
\label{eq:18}
\end{equation}
where $E_W$ is the sum of squares of the network weights (including biases).  The
multipliers $\tilde{\alpha}$ and $\tilde{\beta}$ are hyperparameters defined by 
\begin{equation}
\tilde{\alpha}_k  = \frac{{\gamma _k }}{{2 E_W }} \qquad {\rm and} \qquad \tilde{\beta}_k  = \frac{{{n} - \gamma _k }}{{2 E_D }},
\label{eq:whole}
\end{equation}
where
\begin{equation}
\gamma _k  = W - 2\tilde{\alpha}  \cdot {\rm tr}\left( {\bf{H}}_k \right)^{ - 1}
\label{gamma}
\end{equation}
is the number of parameters (weights and biases) that are being effectively used by the network, ${n}$ is the number of errors, $W$ is the total number of parameters characterizing the network model (See Eq.~(\ref{eq:1})) and ${\bf{H}} =\nabla ^{{2}} {{F}}$ is the Hessian matrix evaluated for the extended (``regularized'') objective function (\ref{eq:18}).  The full Hessian computation is again bypassed using the Gauss-Newton approximation, writing
\begin{equation}
{\bf{H}}_k=\tilde{\beta}_k \nabla ^{{2}} {{E}}_{{D}} {{ + }}\tilde{\alpha}_k \nabla ^{{2}} {{E}}_{{W}}  \approx 2\tilde{\beta}_k {\bf{J}}^{\rm T}_k {\bf{J}}_k + 2\tilde{\alpha}_k {\bf{I}} .
\label{eq:22}
\end{equation}
Thus, the Levenberg-Marquardt optimization algorithm updates the weights/biases by means of the rule
\begin{equation}
{\bf{w}}_{k+1}  = {\bf{w}}_{k}  - \left[ {\tilde{\beta}_k {\bf{J}}^{\rm T}_k {\bf{J}}_k + \left( {\mu_k  + \tilde{\alpha}_k } \right){\bf{I}}} \right]^{ - 1} \left( {\tilde{\beta}_k {\bf{J}}^{\rm T}_k {\bf{e}}_k + \tilde{\alpha}_k {\bf{w}}_{k} } \right).
\label{eq:23}
\end{equation}
A detailed discussion of the use of Bayesian regularization in combination with the Levenberg-Marquardt algorithm can be found in Ref.~\onlinecite{16}.

\subsection{\label{subsubsec:leve26}Training Mode}

Backpropagation learning, as a technique for iterative updating of network parameters, can be executed in either the {\it batch} or {\it pattern-by-pattern} (or ``on-line'') mode.  In  the on-line mode, a pattern is presented to the network and its response recorded; the Jacobian
matrix is then computed and the weights/biases updated {\it before} the next pattern is presented.  In the batch mode, on the other hand, calculation of the Jacobian and parameter updating is performed only after all training examples have been presented, i.e., at the end of each epoch.  The model results reported here are based on the batch mode, the choice being made on the empirical basis of findings from a substantial number of computer experiments carried out with both strategies.

\subsection{\label{subsubsec:leve27}Data Sets}

\begin{figure*}[htb]
\includegraphics [width=6.8in] {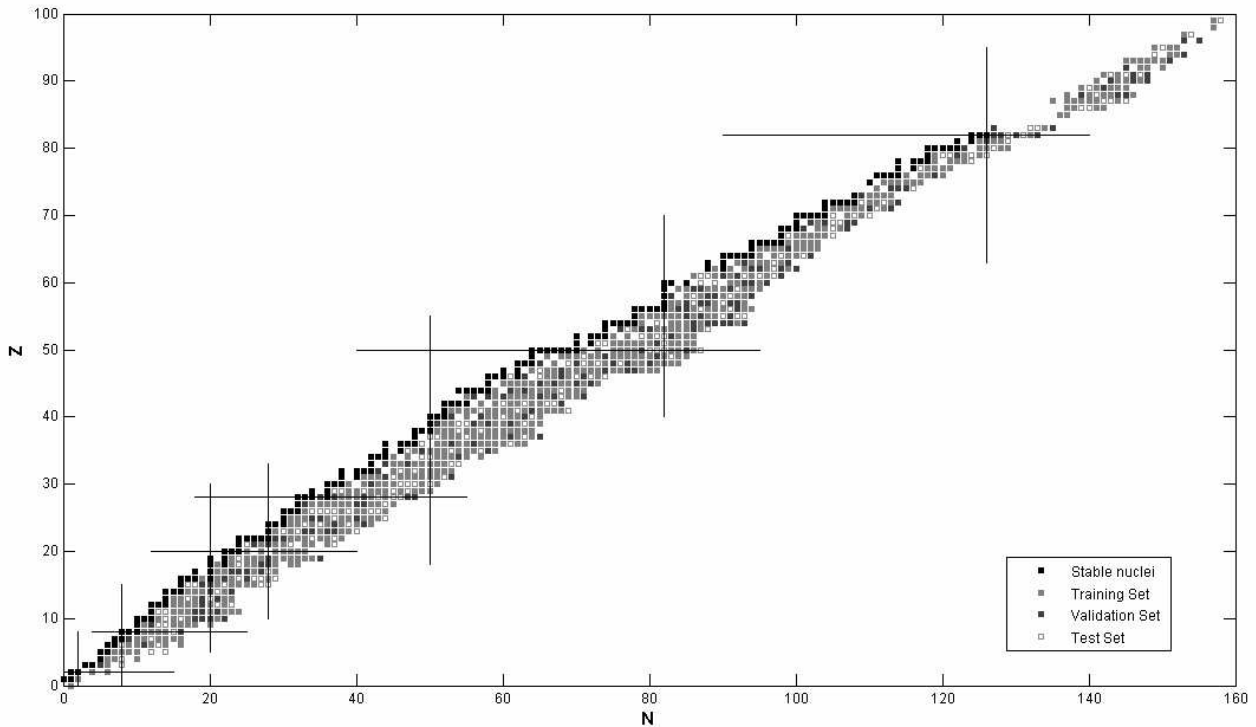}
\caption{\label{fig:3E} The partitioning of the whole set of halflives in the learning, validation, and test sets as a function of the atomic ($Z$) and the neutron ($N$) numbers. Stable nuclides are also indicated.}
\end{figure*}

\begin{figure}[htb]
\includegraphics [width=3.47in] {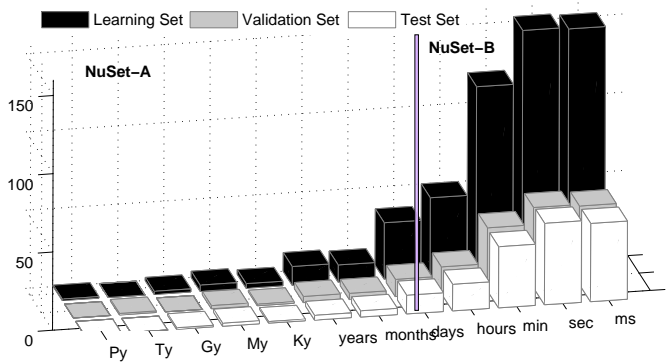}
\caption{\label{fig:1E} Distribution of halflives over the timescale for NuSet-A nuclides. NuSet-B nuclides lie to the right of the vertical gray rectangle.}
\end{figure}

The experimental data used in developing ANN
models of $\beta$-decay systematics have been taken
from the Nubase2003 evaluation \cite{28} of nuclear and decay
properties carried out by Audi et al.\ at the Atomic
Mass Data Center.  Restricting attention to those
cases in which the ground state of the parent decays 100\%
through the $\beta^-$ channel, we form a subset of the
beta-decay data denoted by NuSet-A, consisting of 905
nuclides sorted by halflife.  The halflives of nuclides
in this set range from $0.15 \times 10^{-2}$ s for
$^{35}${\rm Na} to $2.43 \times 10^{23}$ s for $^{113}${\rm Cd}.
Of these NuSet-A nuclides, 543 (60\%) have been 
chosen, at random with a uniform probability,
to form the training
set, and 181 (20\%) of those remaining have been
similarly chosen to form the validation set.  The residual
181 (20\%) are reserved for testing the predictive
capability of the models constructed.  Such partitioning
of the NuSet-A database (uniform selection) was
implemented to ensure that the distribution over halflives
in the whole set is faithfully reflected in the learning,
validation, and test sets. Fig.~2 shows an example of
the results of this procedure, as viewed in the $Z-N$ diagram.

We also formed a more restricted data set, called NuSet-B, by eliminating
from NuSet-A those nuclei having halflife greater than $10^6\,{ s}$. The halflives in this subset, which consists of 838 nuclides, range from $0.15 \times 10^{-2}\,{ s}$ for $^{35}${\rm Na} to $0.20\times 10^{6}\,{ s}$ for $^{247}${\rm Pu}.  
Histograms depicting the lifetime distribution of the NuSet-B nuclides are shown in Fig.~\ref{fig:1E}, having made a uniform subdivision of the data into learning, validation, and test sets, consisting respectively of 503 ($\sim$ 60\%),  167 ($\sim$ 20\%), and 168 ($\sim$ 20\%) examples.  Having excluded the few long-lived examples from NuSet-A (situated to the right of the vertical line in Fig.~3), one is then dealing with a more homogeneous collection of nuclides, a property that facilitates the training of network models.  Accordingly, we have focused our efforts on NuSet-B.  Table~\ref{tab:li} gives information on the distribution of NuSet-B nuclides with respect to the even versus odd character of $Z$ and $N$.

When considering the performance of a network model for examples taken from the whole data set (whether NuSet-A or NuSet-B), we speak of operation in the \textit{Overall Mode}.   Similarly, we speak of operation in the \textit{Learning}, \textit{Validation}, and \textit{Prediction Modes} when studying performance on the learning, validation, and test sets, respectively.

\subsection{\label{subsec:pairing}Coding Schemes at Input and Output Interfaces.}

In our initial experiments in the design of ANN models for $\beta$-decay
halflife prediction, we employed
input coding schemes that involve only the proton number $Z$ and the neutron number $N$. To keep the number of weights to a minimum, we make use of analog (i.e., {\it floating-point})
coding of $Z$ and $N$ through two dedicated inputs, whose activities
represent scaled values of these variables.  
The LMOBP algorithm works better when the network inputs and targets  
are scaled to the interval $[-1,1]$ than (say) the interval $[0,1]$~\cite{19}.
Moreover, the range of the hyperbolic tangent activation function employed by the hidden units lies in the interval $ - 1 \le \varphi \left( {{u}} \right) \le 1$.  The ranges [0,230] and [0,230] of $Z$ and $N$ are therefore scaled to this interval. The base-10 log of the $\beta^-$ halflife $T_{\beta,{\rm calc}}$,
as calculated by the network for input nuclide ($Z_p$,$N_p$), is represented by the activity of a single analog output unit. For the same reason as indicated for the input units, the range [0.17609, 8.9771]
of the target values ${{\log}}_{{{10}}}  {{{T}}_{{{\beta }}{{,{\rm exp}}}}^{{p}} } $ is scaled again to the interval [-1,1].

Also in the primary stages of our study of beta-halflife systematics, we
have assumed that the halflife of a given nucleus is properly given 
by an expression of the form of Eq.~(\ref{eq:6}). 
Such an expression echos the essence of Weizsacker's semi-empirical mass  
formula based on the liquid-drop model, with the binding energy given by a function $B(Z,N)$ representing 
a statistical estimate of the physical quantity, plus an additive noise term.

Taking $Z$ and $N$ as the only inputs to the inference machine formed by the neural network has, of course, the logistical advantage that there is no limitation to the range of prediction of nuclear properties across the nuclear landscape.  If, on the other hand, such quantities as $Q$-values and neutron separation energies were included as inputs, one would have to calculate these quantities for choices of $(Z,N)$ at which experimental values are not available.    But this implies a departure from the ``ideal'' of determining the physical mapping from $(Z,N)$ to the target nuclear property,
based {\it only} on the existing body of experimental data for that property.
The predictions of the network model would necessarily be contingent on
some theoretical model to provide the additional values of the input quantities.

However, estimating a given nuclear property -- the log lifetime of 
beta decay in the present case -- as a smooth function of $Z$ and
$N$ has clear limitations.   The nuclear data itself sends strong 
messages of the importance of pairing and shell effects (``quantal 
effects'') associated with the integral nature of $Z$ and $N$.
The problem of atomic masses provides the classic example:
the liquid drop formula must be supplemented by pairing and
shell corrections to account for the existence of different mass
surfaces for even-even, odd-$A$, and odd-odd nuclei and other effects 
of the integral/particulate character of $Z$ and $N$.

\begin{figure}[htb]
\includegraphics[width=3.4in]{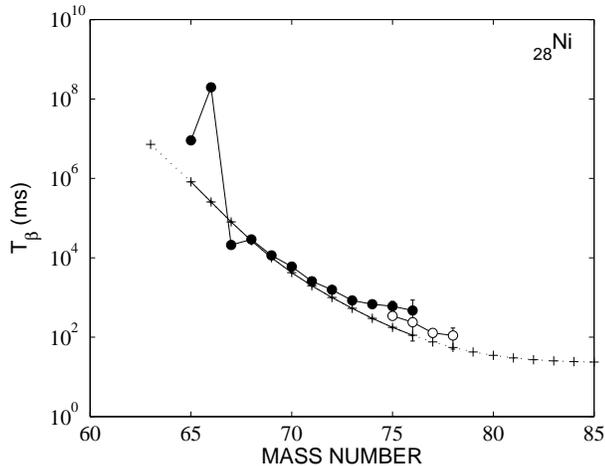}
\caption{\label{fig:fig4} Plot showing calculated and experimental
$\beta^-$-decay halflives for the $_{28}$\rm{Ni} isotopic chain. Solid
dots: experimental data points.  Unfilled dots: new and more precise
experimental halflives recently deduced by Hosmer et al.~\cite{20}.
Pluses: results generated by the
$\left[{ 2-5-5-5-5-1\left|111\right.} \right]$ ANN model with inputs
$(Z,N)$.  Solid lines trace the calculated values of the Overall Mode
(learning, validation, and test sets), while dotted lines trace
extrapolated values produced by the model.}
  
\end{figure}

Examination of results from the simple coding scheme with
$Z$ and $N$ alone serving as analog inputs  is nevertheless 
instructive.   We have applied the LMBP training algorithm to develop a
network model with architecture $\left[{2-5-5-5-5-1\left|111\right.} \right]$. 
As shown in Fig.~\ref{fig:fig4}, the model yields a smooth curve that represents a gross fit of the experimental data involved.  The predictive ability of the model naturally relies on extrapolation based on this curve.  These results demonstrate the need for a more refined model within which quantal effects such as pairing and shell structure are given an opportunity to exert themselves, so that the natural fluctuations are followed in validation and prediction modes, as well as in the learning (or ``fitting'') phase.

A straightforward modification of the input interface of the network model that can at least partially fulfill this need is suggested by the extension of the liquid-drop model to include a pairing-energy term.    In addition to the two input units representing $Z$ and $N$ as floating-point numbers, we introduce a third input unit representing
a discrete parameter analogous to the pairing constant, namely
\begin{eqnarray}
\delta  = \left\{ {\begin{array}{*{20}c}
   { + 1,}  \\
   {0,}  \\
   { - 1,}  \\
\end{array}\begin{array}{*{20}c}
   {\begin{array}{*{20}c}
   {\rm for} & {\rm e - e} & {\rm nuclei},  \\
\end{array}}  \\
   {\begin{array}{*{20}c}
   {\rm for} & {\rm o - mass} & {\rm nuclei},  \\
\end{array}}  \\
   {\begin{array}{*{20}c}
   {\rm for} & {\rm o - o} & {\rm nuclei} , \\
\end{array}}  \\
\end{array}} \right.  
\label{eq:28}
\end{eqnarray}
which distinguishes between even-$Z$-even-$N$, odd-$A$, and odd-$Z$-odd-$N$ nuclides. This simple refinement has the conceptual
advantage of remaining in the spirit of ``theory-thin'' modeling,
driven purely by data rather than data plus physical intuition
and accepted theory.  All that is required is the knowledge that $Z$ and $N$ are actually integers and recognition of their even or odd parity.  The expression replacing Eq.~(\ref{eq:6}) as a representation of the inference process performed by the ANN model is evidently
\begin{equation}
\log_{10} T_\beta(Z,N) = {\tilde g}(Z,N,\delta) + \tilde \varepsilon(Z,N).
\label{eq:226}
\end{equation}

We shall see that some shell effects that might impact the behavior of halflives for both allowed and/or forbidden transitions can, at least to some extent, be taken into account by the $\delta$ input defined in Eq.~(\ref{eq:28}).
It should be mentioned that in the ANN global models of nuclear
mass excess \cite{70}, it has proven advantageous to introduce two
binary input units that encode the even/odd parity of $Z$ and $N$.

\subsection{\label{subsubsec:leve29}Initialization of Network Parameters}

Proper initialization of the free parameters of the ANN -- its weights
and biases -- is a very important and highly nontrivial task.  One needs to choose an initial point on the error surface defined by Eqs.~(\ref{eq:7}),~(\ref{eq:18}) as close as possible to its global minimum with respect to these parameters, and 
such that the output of each neuronal unit lies within the sensitive region of its activation function $\phi$. We adopt a method  devised by  
Nguyen and Widrow \cite{21}, in which the initial weights are  
selected so as to distribute the active region of each neuron
(its ``receptive field'' neurobiological parlance) approximately
evenly across the input space of the layer to which that neuron   
belongs.  The Nguyen-Widrow method has clear advantages over
more naive initializations in that all neurons begin operating with
access to good dynamical range, and all regions of the input space 
receive coverage from neurons.  Consequently, training of the
network is accelerated.

\section{\label{sec:performance}PERFORMANCE MEASURES}

The performance of the models we have been developing is assessed in terms of several commonly used statistical measures, namely, the \textit{Root Mean Square Error} ($\sigma_{\rm RMSE}$), the \textit{Mean Absolute Error} ($\sigma_{\rm MAE}$), and the \textit{Normalized Mean Square Error} ($\sigma_{\rm NMSE}$).  For any given data set, these quantities provide overall measures of  the deviation of the calculated values
$y_i \equiv \log_{10}T_{\beta,{\rm calc}}$ of the log-halflife produced
by the model for nuclide $i$, from the corresponding experimental value $\hat{y}_i \equiv \log_{10}T_{\beta,{\rm exp}}$.   To understand the network's response in more detail, a \textit{Linear Regression Analysis} (\rm{LR}) is also carried out in which the correlation between experimental and calculated halflife values is evaluated in terms of the correlation coefficient (\rm{R}-value).  Definitions of these quantities follow, with $n$ standing for the total number of nuclides in each case (the full data set or one of its subsets --
the learning, validation, or test set).

\noindent
{}

\noindent
\textit{Root Mean Square Error}
\begin{equation}
{\sigma_{\rm RMSE}} = \left[ {\frac{1}{n}\sum_{p=1}^n {\left( {y_p  - \hat y_p } \right)^2 } } \right]^{1/2}.
 \label{eq:29}
\end{equation}
\textit{Normalized Mean Square Error}
\begin{equation}
{\sigma_{\rm NMSE}} = \frac{{\sum_{p=1}^n {\left( {y_p  - \hat y_p } \right)^2 } }}{{\sum_{p=1}^n {\left( {y_p  - \bar y_p } \right)^2 } }} .
\label{eq:30}
\end{equation}
\textit{Mean Absolute Error}
\begin{equation}
{\sigma_{\rm MAE}} = \frac{1}{n}\sum_{p=1}^n {\left| {y_p  - \hat y_p } \right|}.
\label{eq:31}
\end{equation} 
Those models having smaller values of $\sigma_{\rm RMSE}$ and $\sigma_{\rm MAE}$, 
and $\sigma_{\rm NMSE}$ closer to unity, are favored.

\noindent
{}

\noindent
\textit{Linear Regression} (\rm{LR})
\begin{equation}
y_p  = a \hat y_p  + b.
\label{eq:39}
\end{equation}
In linear regression, the slope $a$ and the intercept $b$ are calculated, as well as the correlation coefficient   
\begin{equation}
\rm{R} = \frac{{\sum_{p=1}^n {Y_p \hat Y_p } }}{{\left( {\sum_{p=1}^n {\left( {Y_p  - \left\langle {Y_p } \right\rangle } \right)^2 } \sum_{p=1}^n {\left[ {\hat Y_p  - \left\langle {\hat Y_p } \right\rangle } \right)^2 } } \right]^{{1 \mathord{\left/
 {\vphantom {1 2}} \right.
 \kern-\nulldelimiterspace} 2}} }},
\label{eq:40}
\end{equation}
where $Y_p  = y_p  - \left\langle y \right\rangle$ and $\hat Y_p  = \hat y_p  - \left\langle {\hat y} \right\rangle $. 
Values of \rm{R} greater than $0.8$ indicate strong correlations. 

\noindent
{}

The above indices necessarily provide only gross assessments  
of the quality of our models.  In the literature on global modeling of $\beta^-$ halflives, several additional indices, perhaps more appropriate to the physical context, have been used to analyze  performance. The collaboration led by Klapdor~\cite{eirhnh11,2,5,eirhnh14,6,eirhnh16} has employed the quality measure 
\begin{equation}
\bar{x}_K  = \frac{1}{n}\sum_{p=1}^n {x_p } ,
\label{eq:41}
\end{equation}
wherein 
\begin{eqnarray}
x_p  = \left\{ {\begin{array}{*{20}c}
   {\begin{array}{*{20}c}
   {{{T_{\beta,{\rm exp} } } \mathord{\left/
 {\vphantom {{T_{\beta,{\rm exp} } } {T_{\beta,{\rm calc}} ,}}} \right.
 \kern-\nulldelimiterspace} {T_{\beta,{\rm calc}} ,}}} & {\rm if} & {T_{\beta,{\rm exp} }  \ge T_{\beta,{\rm calc}} }  \\
\end{array}}  \\
   {\begin{array}{*{20}c}
   {{{T_{\beta,{\rm calc}} } \mathord{\left/
 {\vphantom {{T_{\beta,{\rm calc}} } {T_{\beta,{\rm exp} } ,}}} \right.
 \kern-\nulldelimiterspace} {T_{\beta,{\rm exp} } ,}}} & {\rm if} & {T_{\beta,{\rm exp} }  < T_{\beta,{\rm calc}} } , \\
\end{array}}  \\
\end{array}} \right. 
 \label{eq:42}
\end{eqnarray}
along with the corresponding standard deviation ${\bar x}_K$
\begin{equation}
\sigma_K  = \left[ {\frac{1}{n}\sum_{p=1}^n {\left( {x_p - \bar{x}_K} \right)^2 } } \right]^{{1 \mathord{\left/
 {\vphantom {1 2}} \right.
 \kern-\nulldelimiterspace} 2}}.
\label{eq:43}
\end{equation}
Again the sums run over the appropriate set of nuclides. Perfect accuracy is attained when
$\bar{x}_K = 1$ and $\sigma_K = 0$.

In a more incisive assessment, also pursued by Klapdor and coworkers, one calculates the percentage $m$ of nuclides
having measured ground-state halflife $T_{\beta,{\rm exp}}$ within a prescribed range (e.g., not greater than $10^6$ $s$, 60 $s$, or 1 $s$), for which the halflife generated by the model is within a prescribed tolerance factor $f$ (in particular, 2, 5, or 10) of the experimental value.

A measure $M$ similar to ${\bar x}_K$, but defined in terms of $\log_{10}T_{\beta}$ rather than $T_{\beta}$, has been used by M\"{o}ller and collaborators~\cite{7,8}; specifically,
\begin{equation}
M  = \frac{1}{n}\sum_{p=1}^n {r_p }, 
\label{eq:45}
\end{equation}
where $r_p = y_p /{\hat y}_p$.  This quantity gives the average 
position of the points in Fig.~\ref{fig:fig6} for the respective
data sets.  Its associated standard deviation 
\begin{equation}
\sigma_M  = \left[ {\frac{1}{n}\sum_{p=1}^n{\left( {r_p  - M } \right)^2 } } \right]^{{1 \mathord{\left/
 {\vphantom {1 2}} \right.
 \kern-\nulldelimiterspace} 2}}, 
\label{eq:46}
\end{equation}
is also examined, and the ``total'' error of the model for the data set in question is taken to be 
\begin{equation}
\Sigma  = \left[ \frac{1}{n}\sum_{p=1}^n  r_p^2   \right]^{{1 \mathord{\left/
 {\vphantom {1 2}} \right.
 \kern-\nulldelimiterspace} 2}} ,
\label{eq:47}
\end{equation}
which is the same as the $\sigma_{\rm RMSE}$ defined in Eq.~(\ref{eq:29}).  Model quality is also expressed in terms of exponentiated versions of these
last three quantities, namely the mean deviation range
\begin{equation}
M^{(10)}  = 10^{M},
\label{eq:48}
\end{equation}
the mean fluctuation range  
\begin{equation}
\sigma_{\rm M^{(10)}}  = 10^{\sigma_M}, 
\label{eq:49}
\end{equation}
and total error range $\Sigma^{(10)}$:
\begin{equation}
\Sigma^{(10)}  = 10^{\Sigma}.
\label{eq:50}
\end{equation}

Superior models should have  $\Sigma$,  $M$, and $\sigma_M$ near zero, and $M^{(10)}$, $ \sigma_M^{(10)}$, and $\Sigma^{(10)}$ near unity.
Again, in a closer analysis of model capabilities, these indices are evaluated within prescribed halflife domains.

\section{\label{sec:level3}RESULTS AND DISCUSSION}

As already indicated, statistical modeling of $\beta^-$-decay systematics 
is more effective when the range of lifetimes considered is more restricted.
Accordingly, the following detailed presentation and analysis will focus on the properties  and performance of the best ANN model developed using the NuSet-B database,
which is restricted to nuclides with $\beta^-$ halflife below $10^6$ s.   The quality  
of this model will be compared, in considerable detail, with that of traditional 
theoretical global models cited in the introduction, earlier ANN models,
and models provided by another class of learning machines (Support Vector
Machines, or SVMs).

After a large number of computer experiments on networks developed with different architectures, input/output coding schemes, activation functions, initialization prescriptions, and training algorithms \cite{Nick}, we have arrived at an ANN model well suited to approximate reproduction of the observed $\beta^-$-decay halflife systematics and prediction of halfives of nuclides unfamiliar to the network.  The preferred network is of architecture $\left[{3-5-5-5-5-1\left|116\right.} \right]$.  The hyperbolic tangent sigmoid is taken as the activation function of neurons in hidden layers, and a saturated linear function
is adopted in the output layer.   In training, the techniques for improving generalization that were described in Sec.~II, namely, Bayesian regularization and cross-validation, were implemented in combination with the Levenberg-Marquardt optimization algorithm (LMOBP) and the Nguyen-Widrow initialization method.
The network was taught in batch mode and the training phase was continued for 696 epochs. 
Of the 116 degrees of freedom corresponding to the network weights and biases,
98 survive the training process; this is the value of the number 
$\gamma_k$ defined in Eq.~(\ref{gamma}).

\subsection{\label{sec:leve31}Comparison with Experiment}

In this subsection, we evaluate the performance of our ANN model by  direct comparison with the available experimental data. Table~\ref{tab:results} collects results for the overall quality measures 
(\ref{eq:29})--(\ref{eq:31}) commonly used in statistical analysis as well as the values of the correlation coefficient \rm{R} (See Eq.~(\ref{eq:40})).We may quote for comparison the root-mean-square errors of 1.08 (learning mode) and 1.82 (prediction mode) obtained in an earlier ANN model
of beta-decay systematics \cite{13}.
 \begin{table}[h]
\caption{\label{tab:results}Performance measures for the learning, validation, test, and whole sets, achieved by the favored ANN model of $\beta^-$-decay halflives, a network with architecture $\left[{3-5-5-5-5-1\left|116\right.} \right]$ trained on nuclides from NuSet-B.}
\begin{ruledtabular}
\begin{tabular}{ccccc}
Performance	&Learning &Validation &Test &Whole \\
Measure &Set &Set &Set &Set\\
\hline
$\sigma_{\rm RMSE}$	&0.53	&0.60	&0.65	&{0.57} \\
$\sigma_{\rm NMSE}$	&1.004	&0.995	&1.012	&{0.999} \\
$\sigma_{\rm MAE}$	     &0.38	&0.41	&0.46	&{0.40} \\
\rm{R}-value	&0.964	&0.953	&0.947 	&{0.958} \\
\end{tabular}
\end{ruledtabular}
\end{table}

\begin{figure}
\includegraphics[width=3.4in]{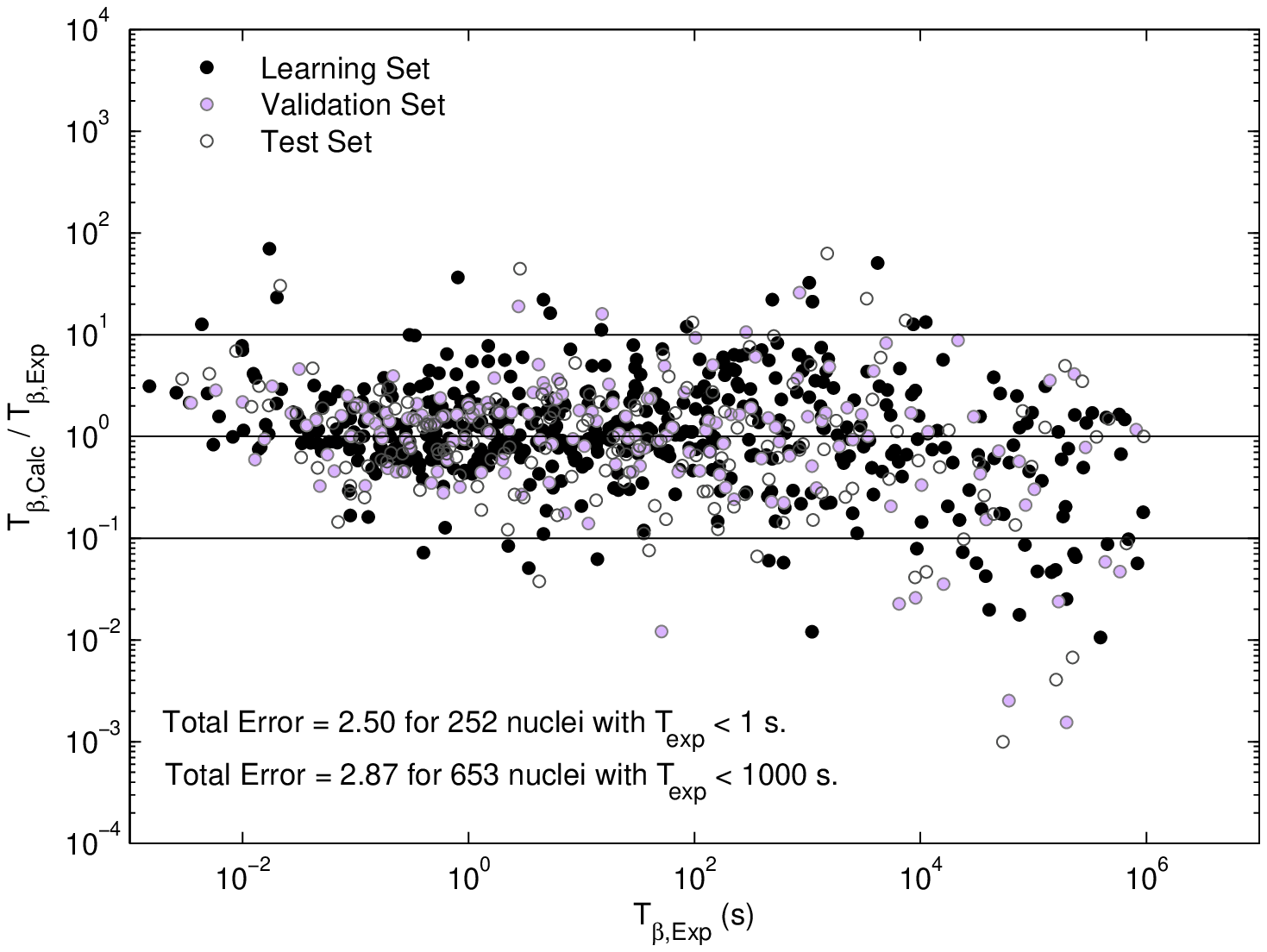}
\caption{\label{fig:fig6} Ratios of calculated to experimental halflives values for nuclides in the learning (black), validation (gray), and test (white) sets selected from NuSet-B, plotted versus halflife $T_{\beta,{\rm exp}}$. Total Error equals to $\Sigma^{(10)}$ (See Eq.~\ref{eq:50}).}

\noindent{}

\noindent{}
 
\includegraphics[width=3.4in]{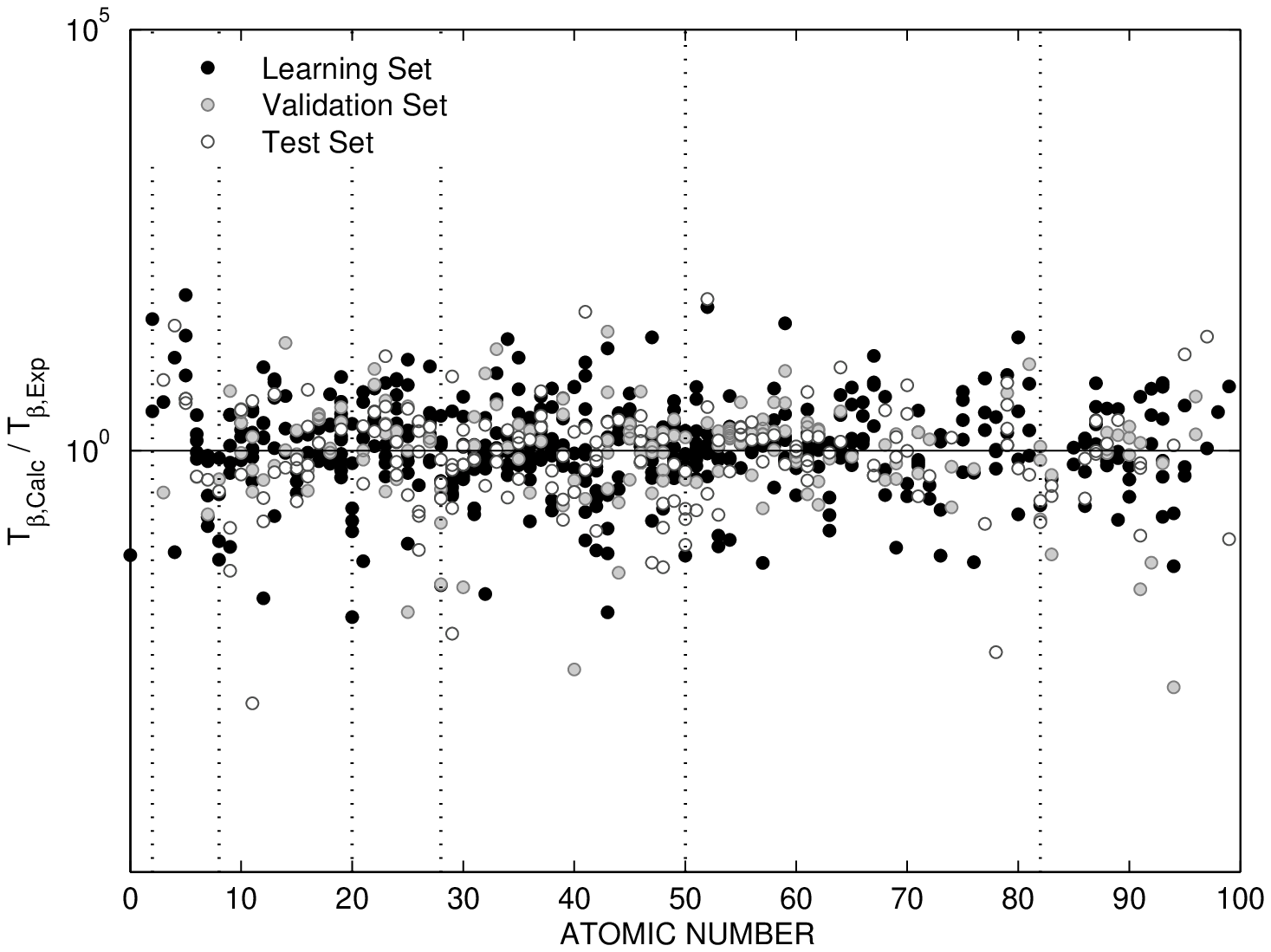}
\caption{\label{fig:fig7} Same as Fig.~\ref{fig:fig6}, but ratios of calculated to experimental halflives are plotted against the atomic number $Z$. The dashed lines indicate the magic numbers. }        
\end{figure}

\begin{figure*}
\includegraphics[width=7in]{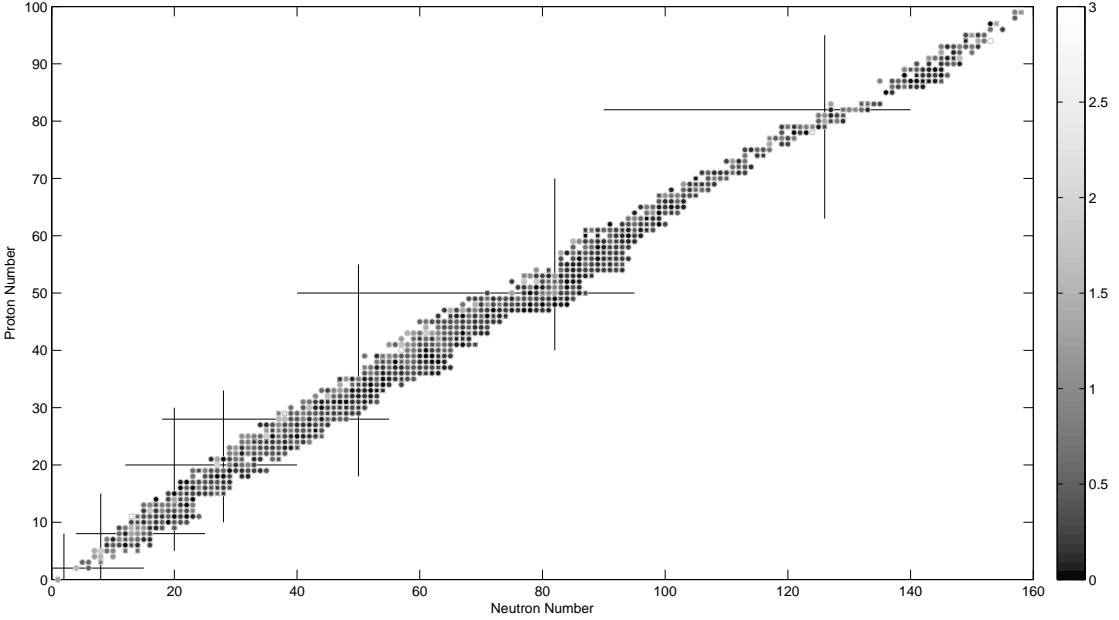}
\caption{\label{fig:fig10} Absolute errors of the calculated to experimental beta-decay halflives of all nuclides $(p)$ in the full NuSet-B database, plotted versus proton and neutron numbers $Z$ and $N$ for the $\left[{3-5-5-5-5-1\left|116\right.} \right]$ network model. The bar on the right indicates the mapping from the absolute error values $|e_p|=|{{ {{{\log}}_{{{10}}} {{{T}}_{{{\beta }}{{,{\rm exp}}}}^{{p}} }{{ - \log}}_{{{10}}} {{{T}}_{{{\beta }}{{,{\rm calc}}}}^{{p}} }}}} |$ to the gray scale. Test nuclides are indicated as squares.}
\end{figure*}

\begin{figure}[b]
\includegraphics[width=3.4in]{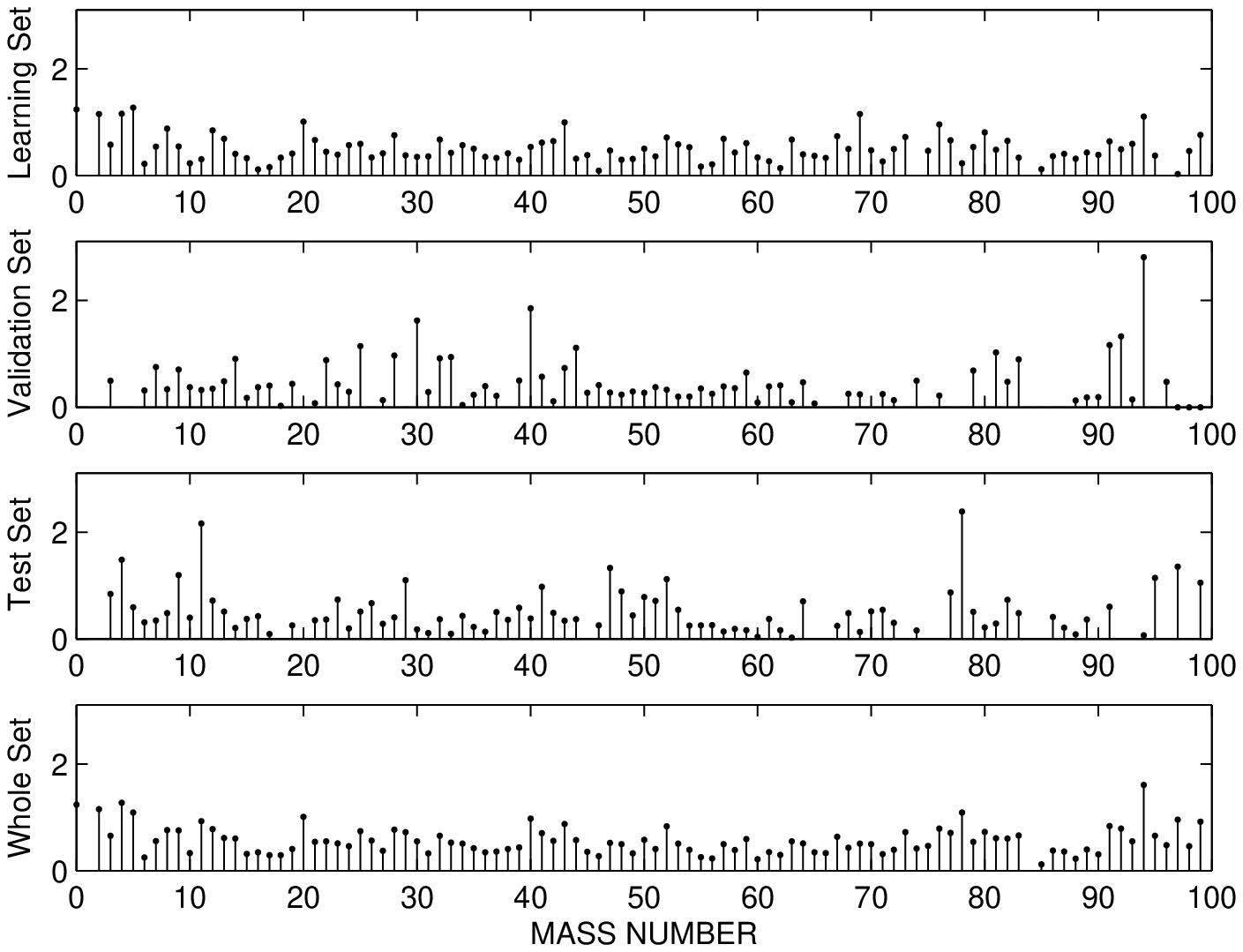}
\caption{\label{fig:zrmse} $\sigma_{\rm RMSE}$ values in each isotopic chain, for the nuclides in the learning, validation, and test sets, and the full NuSet-B database, plotted against the mass number $A$, for the   
$\left[{3-5-5-5-5-1\left|116\right.} \right]$ network model.}
\end{figure}

\begin{figure*}

\includegraphics[width=6in]{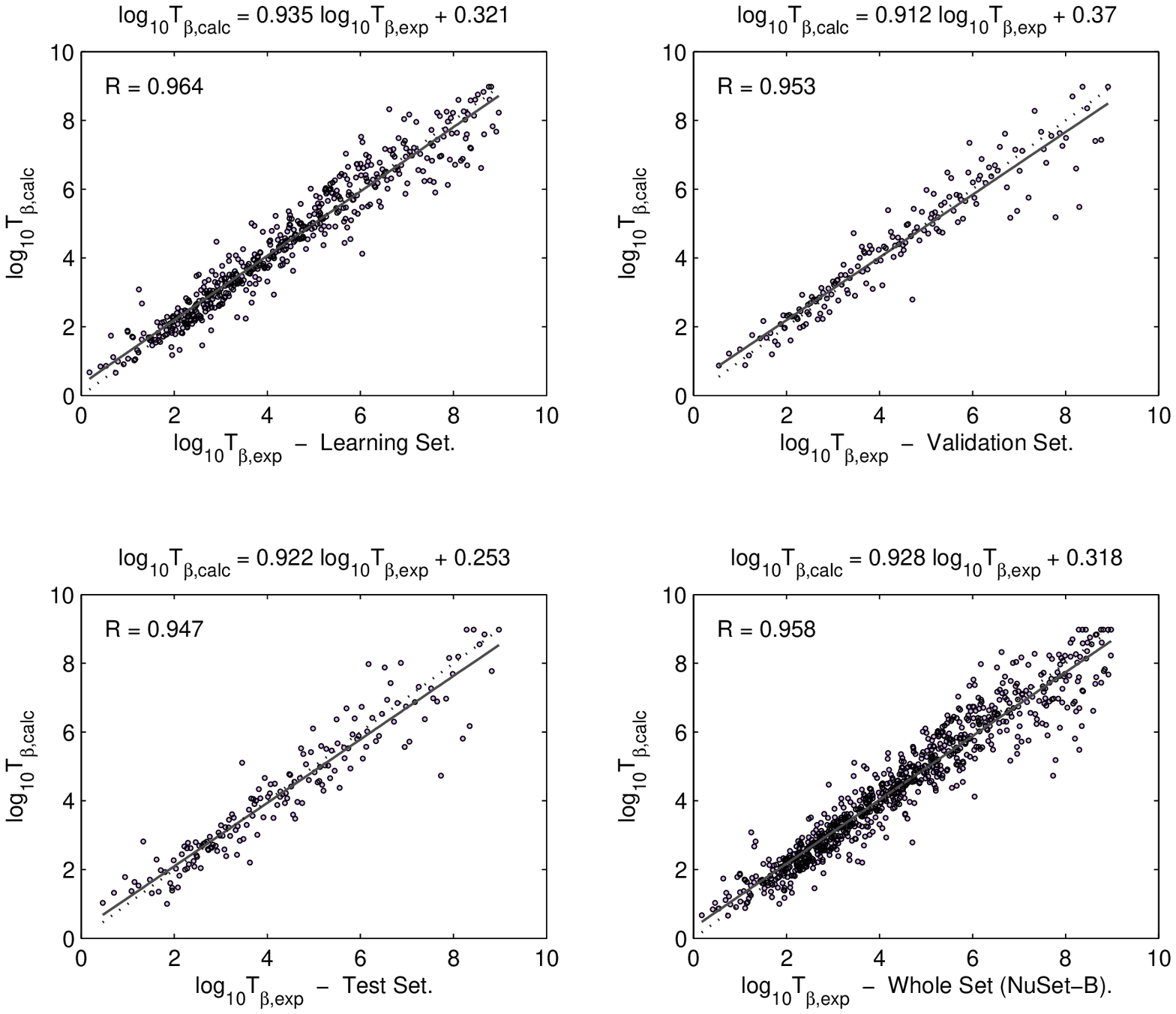}
\caption{\label{fig:fig9} Regression analysis for the learning, validation, test (prediction mode) and for the full
database (overall mode). Solid lines represent the desirable relation: $\left( \log_{10}  T_{\beta,{\rm calc}}  =\log_{10} T_{\beta,{\rm exp}} \right)  $, while dashed lines indicate the corresponding best linear fittings. The respective values of the parameters $a$ and $b$ of Eq.~(\ref{eq:39}) 
and the correlation coefficient \rm{R} of (Eq.~(\ref{eq:40})) are
given in each panel.}
\end{figure*}

These overall measures are
silent with respect to specific physical merits or shortcomings of the model.  On the other hand, such information can be revealed by suitable plots of the results from applications
of the model, as exemplified in Figs.~\ref{fig:fig6}--\ref{fig:fig9}.

Figs.~\ref{fig:fig6} and~\ref{fig:fig7} present the ratios of calculated to experimental halflife values. The deviations from the measured values are clearly visible as departures from the solid line ${{{{T}}_{{{\beta }}{{,{\rm calc}}}} } \mathord{\left/
 {\vphantom {{{{T}}_{{{\beta }}{{,{\rm calc}}}} } {{{T}}_{{{\beta }}{{,{\rm exp}}}} }}} \right.
 \kern-\nulldelimiterspace} {{{T}}_{{{\beta }}{{,{\rm exp}}}} }} = 1
$. Both figures show that the model response follows the general trend of experimental halflives. The scattered points at higher halflife values imply that forbidden transitions are not adequately taken into account by the model. On the other hand, shell effects are included in the right direction as shown in Figs.~\ref{fig:fig7}--\ref{fig:zrmse}. The accuracy of model output versus distance from stability can be inferred from Fig.~\ref{fig:fig10}.  The local isotopic $\sigma_{\rm RMSE}$ (Fig.~\ref{fig:zrmse}) 
and the absolute deviation of calculated from experimental $\log_{10}T_\beta$ values (Fig.~\ref{fig:fig10}) indicate a balanced behavior of network response in all $\beta^-$-decay regions. However, Fig.~\ref{fig:fig10} shows that some less accurate results are obtained very near the $\beta$-stability line, a feature also present in the traditional models of Refs.~\onlinecite{8,6}. For nuclei with very small or very large mass values there are no
significant deviations.

Finally, the regression analysis we have performed,  in which
linear fits are made for the learning, validation, and test sets as well
as the full NuSet-B database, serves to demonstrate in a different 
way the slight discrepancies between calculated and observed $\beta^-$-decay halflives, as illustrated in Fig.~\ref{fig:fig9}.  Moreover, the resultant \rm{R}-values (See also Table~\ref{tab:results}) imply that the observed systematics is smoothly and uniformly mirrored in the model's responses.

\begin{table}
\caption{\label{tab:odd-mass-pres}
Analysis of 
the deviation between calculated and experimental $\beta^-$-decay halflives  of the
$\left[{3-5-5-5-5-1\left|116\right.} \right]$ 
ANN model in the Overall and Prediction Modes, based on the quality measures $M^{(10)}$ and $\sigma_{\rm M^{(10)}}$ of
Eqs.~(\ref{eq:48})--(\ref{eq:49}) used by M\"oller and coworkers.  The second column 
denotes the even/odd character of the parent nucleus in $Z$ and $N$, while $n$ is the number of nuclides with experimental halflives lying in the prescribed range (first column).}
\begin{ruledtabular}
\begin{tabular}{lllll}
\multicolumn{1}{c}{$ T_{\beta,{\rm exp}}$} & \multicolumn{4}{r}{ (a) ANN Model. Overall Mode.}\\ 
\multicolumn{1}{c}{(s)} & \multicolumn{1}{c}{Class} &  \multicolumn{1}{c}{$n$} & \multicolumn{1}{c}{$M^{(10)}$} & \multicolumn{1}{c}{$\sigma_{\rm M^{(10)}}$}  \\ 
\hline
\multicolumn{1}{l}{$<1$} & \multicolumn{1}{c}{o-o} & \multicolumn{1}{c}{76} & \multicolumn{1}{c}{1.04} & \multicolumn{1}{c}{2.53} \\ 
\multicolumn{1}{c}{ } & \multicolumn{1}{c}{odd} & \multicolumn{1}{c}{125} &\multicolumn{1}{c}{1.16} & \multicolumn{1}{c}{2.25} \\ 
\multicolumn{1}{c}{ } &\multicolumn{1}{c}{e-e} & \multicolumn{1}{c}{51} &\multicolumn{1}{c}{1.87} & \multicolumn{1}{c}{2.45} \\ 
\multicolumn{1}{c}{ } & \multicolumn{1}{c}{} & \multicolumn{1}{c}{} &\multicolumn{1}{c}{} & \multicolumn{1}{c}{} \\ 
\multicolumn{1}{l}{$<10$} & \multicolumn{1}{c}{o-o} & \multicolumn{1}{c}{121} & \multicolumn{1}{c}{1.11} & \multicolumn{1}{c}{2.96} \\ 
\multicolumn{1}{c}{ } & \multicolumn{1}{c}{odd} & \multicolumn{1}{c}{187} &\multicolumn{1}{c}{1.10} & \multicolumn{1}{c}{2.31} \\ 
\multicolumn{1}{c}{ } & \multicolumn{1}{c}{e-e} & \multicolumn{1}{c}{87} &\multicolumn{1}{c}{1.65} & \multicolumn{1}{c}{2.56} \\ 
\multicolumn{1}{c}{ } & \multicolumn{1}{c}{} & \multicolumn{1}{c}{} &\multicolumn{1}{c}{} & \multicolumn{1}{c}{} \\ 
\multicolumn{1}{l}{$<100$} & \multicolumn{1}{c}{o-o} & \multicolumn{1}{c}{158} &\multicolumn{1}{c}{1.08} & \multicolumn{1}{c}{3.06} \\ 
\multicolumn{1}{c}{ } & \multicolumn{1}{c}{odd} & \multicolumn{1}{c}{261} &\multicolumn{1}{c}{1.08} & \multicolumn{1}{c}{2.45} \\ 
\multicolumn{1}{c}{ } & \multicolumn{1}{c}{e-e}  & \multicolumn{1}{c}{110} &\multicolumn{1}{c}{1.58} & \multicolumn{1}{c}{2.31} \\ 
\multicolumn{1}{c}{ } & \multicolumn{1}{c}{} & \multicolumn{1}{c}{} &\multicolumn{1}{c}{} & \multicolumn{1}{c}{} \\ 
\multicolumn{1}{l}{$<1000$} & \multicolumn{1}{c}{o-o} & \multicolumn{1}{c}{191} &\multicolumn{1}{c}{1.12} & \multicolumn{1}{c}{3.06} \\ 
\multicolumn{1}{c}{ } & \multicolumn{1}{c}{odd} & \multicolumn{1}{c}{329} &\multicolumn{1}{c}{1.07} & \multicolumn{1}{c}{2.73} \\ 
\multicolumn{1}{c}{ } & \multicolumn{1}{c}{e-e} & \multicolumn{1}{c}{133} &\multicolumn{1}{c}{1.63} & \multicolumn{1}{c}{2.60} \\ 
\multicolumn{1}{c}{ } & \multicolumn{1}{c}{} & \multicolumn{1}{c}{} &\multicolumn{1}{c}{} & \multicolumn{1}{c}{} \\ 
\multicolumn{1}{l}{$<10^6$} & \multicolumn{1}{c}{o-o} & \multicolumn{1}{c}{238} & \multicolumn{1}{c}{0.93} & \multicolumn{1}{c}{3.87} \\ 
\multicolumn{1}{c}{ } & \multicolumn{1}{c}{odd} & \multicolumn{1}{c}{437} &\multicolumn{1}{c}{0.97} & \multicolumn{1}{c}{3.67} \\ 
\multicolumn{1}{c}{ } & \multicolumn{1}{c}{e-e} & \multicolumn{1}{c}{163} &\multicolumn{1}{c}{1.25} & \multicolumn{1}{c}{3.44} \\ 
\hline
\multicolumn{1}{c}{ } & \multicolumn{1}{c}{} & \multicolumn{1}{c}{} & \multicolumn{1}{c}{} & \multicolumn{1}{c}{} \\ 
\hline
\multicolumn{1}{c}{$ T_{\beta,{\rm exp}}$} & \multicolumn{4}{r}{ (b) ANN Model. Prediction Mode.}\\ 
\multicolumn{1}{c}{(s)} & \multicolumn{1}{c}{Class} &  \multicolumn{1}{c}{$n$} & \multicolumn{1}{c}{$M^{(10)}$} & \multicolumn{1}{c}{$\sigma_{\rm M^{(10)}}$}\\  
\hline
\multicolumn{1}{l}{$<1$} & \multicolumn{1}{c}{o-o} & \multicolumn{1}{c}{11} &\multicolumn{1}{c}{0.86} & \multicolumn{1}{c}{1.98} \\ 
\multicolumn{1}{c}{ } & \multicolumn{1}{c}{odd} & \multicolumn{1}{c}{32} &\multicolumn{1}{c}{1.05} & \multicolumn{1}{c}{2.40} \\ 
\multicolumn{1}{c}{ } &\multicolumn{1}{c}{e-e} & \multicolumn{1}{c}{7} &\multicolumn{1}{c}{2.36} & \multicolumn{1}{c}{3.26} \\ 
\multicolumn{1}{c}{ } & \multicolumn{1}{c}{} & \multicolumn{1}{c}{} &\multicolumn{1}{c}{} & \multicolumn{1}{c}{} \\ 
\multicolumn{1}{l}{$<10$} & \multicolumn{1}{c}{o-o} & \multicolumn{1}{c}{20} &\multicolumn{1}{c}{0.86} & \multicolumn{1}{c}{3.76} \\ 
\multicolumn{1}{c}{ } & \multicolumn{1}{c}{odd} & \multicolumn{1}{c}{42} & \multicolumn{1}{c}{0.92} & \multicolumn{1}{c}{2.61} \\ 
\multicolumn{1}{c}{ } & \multicolumn{1}{c}{e-e} & \multicolumn{1}{c}{17} &\multicolumn{1}{c}{1.80} & \multicolumn{1}{c}{2.58} \\ 
\multicolumn{1}{c}{ } & \multicolumn{1}{c}{} & \multicolumn{1}{c}{} & \multicolumn{1}{c}{} & \multicolumn{1}{c}{} \\ 
\multicolumn{1}{l}{$<100$} & \multicolumn{1}{c}{o-o} & \multicolumn{1}{c}{28} &\multicolumn{1}{c}{0.76} & \multicolumn{1}{c}{3.20} \\ 
\multicolumn{1}{c}{ } & \multicolumn{1}{c}{odd} & \multicolumn{1}{c}{57} &\multicolumn{1}{c}{0.97} & \multicolumn{1}{c}{2.91} \\ 
\multicolumn{1}{c}{ } & \multicolumn{1}{c}{e-e}  & \multicolumn{1}{c}{21} & \multicolumn{1}{c}{1.58} & \multicolumn{1}{c}{2.98} \\ 
\multicolumn{1}{c}{ } & \multicolumn{1}{c}{} &\multicolumn{1}{c}{} & \multicolumn{1}{c}{} & \multicolumn{1}{c}{} \\ 
\multicolumn{1}{l}{$<1000$} & \multicolumn{1}{c}{o-o} & \multicolumn{1}{c}{35} &\multicolumn{1}{c}{0.78} & \multicolumn{1}{c}{3.13} \\ 
\multicolumn{1}{c}{ } & \multicolumn{1}{c}{odd} & \multicolumn{1}{c}{68} &\multicolumn{1}{c}{0.84} & \multicolumn{1}{c}{3.07} \\ 
\multicolumn{1}{c}{ } & \multicolumn{1}{c}{e-e} & \multicolumn{1}{c}{28} &\multicolumn{1}{c}{1.49} & \multicolumn{1}{c}{3.04} \\ 
\multicolumn{1}{c}{ } & \multicolumn{1}{c}{} & \multicolumn{1}{c}{} &\multicolumn{1}{c}{} & \multicolumn{1}{c}{} \\ 
\multicolumn{1}{l}{$<10^6$} & \multicolumn{1}{c}{o-o} &\multicolumn{1}{c}{46} & \multicolumn{1}{c}{0.58} & \multicolumn{1}{c}{4.71} \\ 
\multicolumn{1}{c}{ } & \multicolumn{1}{c}{odd} & \multicolumn{1}{c}{87} &\multicolumn{1}{c}{0.86} & \multicolumn{1}{c}{4.07} \\ 
\multicolumn{1}{c}{ } & \multicolumn{1}{c}{e-e} & \multicolumn{1}{c}{35} &\multicolumn{1}{c}{1.14} & \multicolumn{1}{c}{4.33} \\ 

\end{tabular}
\end{ruledtabular}
\end{table}

\begin{table}
\caption{\label{tab:odd-mass-others} 
Same analysis as presented in Table~\ref{tab:odd-mass-pres}, but instead assessing the quality of
traditional theoretical models, corresponding specifically to (a) the NBCS+$pn$QRPA calculation of Homma et al.~\cite{6}, (b) the FRDM+$pn$QRPA calculation of M\"{o}ller and coworkers~\cite{7}, and (c) the SGT calculation by Nakata et 
al.~\cite{12}.   Also, these assessments are
limited to nuclides with experimental halflives below 1000 s.}
\begin{ruledtabular}
\begin{tabular}{lllll}
\multicolumn{1}{c}{$ T_{\beta,{\rm exp}}$} & \multicolumn{4}{c}{ (a) NBCS+$pn$QRPA Calculation~\cite{6}.} \\
\multicolumn{1}{c}{(s)} & \multicolumn{1}{c}{Class} &  \multicolumn{1}{c}{$n$} & \multicolumn{1}{c}{$M^{(10)}$} & \multicolumn{1}{c}{$\sigma_{\rm M^{(10)}}$}  \\ 
\hline
\multicolumn{1}{l}{$<1$} & \multicolumn{1}{c}{o-o} & \multicolumn{1}{c}{28} & \multicolumn{1}{c}{1.75} & \multicolumn{1}{c}{4.96} \\ 
\multicolumn{1}{c}{} & \multicolumn{1}{c}{odd} & \multicolumn{1}{c}{31} & \multicolumn{1}{c}{0.60} & \multicolumn{1}{c}{2.24} \\ 
 \multicolumn{1}{c}{} & \multicolumn{1}{c}{e-e} & \multicolumn{1}{c}{10} & \multicolumn{1}{c}{1.15} & \multicolumn{1}{c}{2.36} \\ 
\multicolumn{1}{c}{ } & \multicolumn{1}{c}{ } & \multicolumn{1}{c}{ } & \multicolumn{1}{c}{ }\\ 
\multicolumn{1}{l}{$<10$} & \multicolumn{1}{c}{o-o} & \multicolumn{1}{c}{66} & \multicolumn{1}{c}{1.89} & \multicolumn{1}{c}{4.60}\\ 
\multicolumn{1}{c}{} & \multicolumn{1}{c}{odd} & \multicolumn{1}{c}{81} & \multicolumn{1}{c}{0.92} & \multicolumn{1}{c}{3.84}\\ 
\multicolumn{1}{c}{ } & \multicolumn{1}{c}{e-e} & \multicolumn{1}{c}{34} & \multicolumn{1}{c}{1.01} & \multicolumn{1}{c}{2.93} \\ 
& \multicolumn{1}{c}{ }\\ 
\multicolumn{1}{l}{$<100$} & \multicolumn{1}{c}{o-o} & \multicolumn{1}{c}{85} & \multicolumn{1}{c}{3.15} & \multicolumn{1}{c}{10.51} \\ 
\multicolumn{1}{c}{ } & \multicolumn{1}{c}{odd} & \multicolumn{1}{c}{127} & \multicolumn{1}{c}{1.07} & \multicolumn{1}{c}{4.29}\\ 
\multicolumn{1}{c}{ } & \multicolumn{1}{c}{e-e} & \multicolumn{1}{c}{52} & \multicolumn{1}{c}{1.13} & \multicolumn{1}{c}{3.58} \\ 
& \multicolumn{1}{c}{ }\\ 
\multicolumn{1}{l}{$<1000$} & \multicolumn{1}{c}{o-o} & \multicolumn{1}{c}{93} & \multicolumn{1}{c}{3.02} & \multicolumn{1}{c}{10.25}\\ 
\multicolumn{1}{c}{ } & \multicolumn{1}{c}{odd} & \multicolumn{1}{c}{157} & \multicolumn{1}{c}{1.10} & \multicolumn{1}{c}{5.55} \\ 
\multicolumn{1}{c}{ } & \multicolumn{1}{c}{e-e} & \multicolumn{1}{c}{63} & \multicolumn{1}{c}{1.39} & \multicolumn{1}{c}{6.10}\\ 
\hline
& \multicolumn{1}{c}{ }\\ 
\hline
\multicolumn{1}{c}{$ T_{\beta,{\rm exp}}$} & \multicolumn{4}{c}{ (b) FRDM+$pn$QRPA Calculation~\cite{7}.}\\ 
\multicolumn{1}{c}{(s)} & \multicolumn{1}{c}{Class} &  \multicolumn{1}{c}{$n$} & \multicolumn{1}{c}{$M^{(10)}$} & \multicolumn{1}{c}{$\sigma_{\rm M^{(10)}}$}  \\
\hline
\multicolumn{1}{l}{$<1$} & \multicolumn{1}{c}{o-o} & \multicolumn{1}{c}{29} & \multicolumn{1}{c}{0.59} & \multicolumn{1}{c}{2.91} \\ 
\multicolumn{1}{c}{ } & \multicolumn{1}{c}{odd} &  \multicolumn{1}{c}{35} & \multicolumn{1}{c}{0.59} & \multicolumn{1}{c}{2.64}\\ 
\multicolumn{1}{c}{} & \multicolumn{1}{c}{e-e} & \multicolumn{1}{c}{10} & \multicolumn{1}{c}{3.84} & \multicolumn{1}{c}{3.08} \\ 
& \multicolumn{1}{c}{ } \\ 
\multicolumn{1}{l}{$<10$} & \multicolumn{1}{c}{o-o} & \multicolumn{1}{c}{59} & \multicolumn{1}{c}{0.76} & \multicolumn{1}{c}{8.83}\\ 
\multicolumn{1}{c}{ } & \multicolumn{1}{c}{odd} & \multicolumn{1}{c}{85} & \multicolumn{1}{c}{0.78} & \multicolumn{1}{c}{4.81}\\ 
\multicolumn{1}{c}{ } & \multicolumn{1}{c}{e-e} & \multicolumn{1}{c}{34} & \multicolumn{1}{c}{2.50} & \multicolumn{1}{c}{4.13}\\ 
& \multicolumn{1}{c}{ } \\ 
\multicolumn{1}{l}{$<100$} & \multicolumn{1}{c}{o-o} & \multicolumn{1}{c}{88} & \multicolumn{1}{c}{2.33} & \multicolumn{1}{c}{49.19} \\ 
\multicolumn{1}{c}{ } &\multicolumn{1}{c}{odd} & \multicolumn{1}{c}{133} & \multicolumn{1}{c}{1.11} & \multicolumn{1}{c}{9.45} \\ 
\multicolumn{1}{c}{ } & \multicolumn{1}{c}{e-e} & \multicolumn{1}{c}{54} & \multicolumn{1}{c}{2.61} & \multicolumn{1}{c}{4.75}\\ 
& \multicolumn{1}{c}{ } \\ 
\multicolumn{1}{l}{$<1000$} & \multicolumn{1}{c}{o-o} & \multicolumn{1}{c}{115} & \multicolumn{1}{c}{3.50} & \multicolumn{1}{c}{72.02} \\ 
\multicolumn{1}{c}{ } & \multicolumn{1}{c}{odd} & \multicolumn{1}{c}{194} & \multicolumn{1}{c}{2.77} & \multicolumn{1}{c}{71.50} \\ 
\multicolumn{1}{c}{ } & \multicolumn{1}{c}{e-e} & \multicolumn{1}{c}{71} & \multicolumn{1}{c}{6.86} & \multicolumn{1}{c}{58.48} \\ 
\hline
& \multicolumn{1}{c}{ }\\ 
\hline
\multicolumn{1}{c}{$ T_{\beta,{\rm exp}}$} & \multicolumn{4}{c}{ (c) SGT Calculation~\cite{12}.} \\ 
\multicolumn{1}{c}{(s)} & \multicolumn{1}{c}{Class} &  \multicolumn{1}{c}{$n$} & \multicolumn{1}{c}{$M^{(10)}$} & \multicolumn{1}{c}{$\sigma_{\rm M^{(10)}}$} \\ 
\hline
\multicolumn{1}{c}{$<1$} & \multicolumn{1}{c}{o-o} &  \multicolumn{1}{c}{38} & \multicolumn{1}{c}{1.45} & \multicolumn{1}{c}{2.57} \\ 
\multicolumn{1}{c}{ } & \multicolumn{1}{c}{odd} & \multicolumn{1}{c}{56} & \multicolumn{1}{c}{1.75} & \multicolumn{1}{c}{2.32} \\ 
\multicolumn{1}{c}{ } & \multicolumn{1}{c}{e-e} & \multicolumn{1}{c}{19} & \multicolumn{1}{c}{2.03} & \multicolumn{1}{c}{2.30} \\ 
& \multicolumn{1}{c}{ } \\ 
\multicolumn{1}{c}{$<10$} & \multicolumn{1}{c}{o-o} & \multicolumn{1}{c}{83} & \multicolumn{1}{c}{1.94} & \multicolumn{1}{c}{4.10} \\ 
\multicolumn{1}{c}{ } & \multicolumn{1}{c}{odd} & \multicolumn{1}{c}{110} & \multicolumn{1}{c}{1.71} & \multicolumn{1}{c}{2.36} \\ 
\multicolumn{1}{c}{ } & \multicolumn{1}{c}{e-e} & \multicolumn{1}{c}{45} & \multicolumn{1}{c}{1.58} & \multicolumn{1}{c}{2.23} \\ 
\multicolumn{1}{c}{ } & \multicolumn{1}{c}{} & \multicolumn{1}{c}{ } & \multicolumn{1}{c}{ } & \multicolumn{1}{c}{ }\\ 
\multicolumn{1}{c}{$<100$} & \multicolumn{1}{c}{o-o} &  \multicolumn{1}{c}{115} & \multicolumn{1}{c}{2.54} & \multicolumn{1}{c}{8.86} \\ 
\multicolumn{1}{c}{ } & \multicolumn{1}{c}{odd} & \multicolumn{1}{c}{174} & \multicolumn{1}{c}{1.95} & \multicolumn{1}{c}{3.15} \\ 
\multicolumn{1}{c}{ } & \multicolumn{1}{c}{e-e} & \multicolumn{1}{c}{64} & \multicolumn{1}{c}{1.45} & \multicolumn{1}{c}{2.40} \\ 
\multicolumn{1}{c}{ } & \multicolumn{1}{c}{} & \multicolumn{1}{c}{ } & \multicolumn{1}{c}{ } & \multicolumn{1}{c}{ }\\ 
\multicolumn{1}{c}{$<1000$} & \multicolumn{1}{c}{o-o} & \multicolumn{1}{c}{144} & \multicolumn{1}{c}{3.42} & \multicolumn{1}{c}{15.21} \\ 
\multicolumn{1}{c}{ } & \multicolumn{1}{c}{odd} & \multicolumn{1}{c}{232} & \multicolumn{1}{c}{2.36} & \multicolumn{1}{c}{5.42} \\ 
\multicolumn{1}{c}{ } & \multicolumn{1}{c}{e-e} & \multicolumn{1}{c}{85} & \multicolumn{1}{c}{1.38} & \multicolumn{1}{c}{2.81} \\ 
\end{tabular}
\end{ruledtabular}
\end{table}

 \begin{table}
\caption{\label{tab:moller}.  Comparison of values of quality indices
characterizing the  ``theory-thin'' neural-network model of the present
work and two ``theory-thick'' models developed by M\"oller and coworkers:
ANN model in Overall (a) and Prediction (b) Modes, and (c) FRDM+$pn$QRPA
and (d) pnQRPA+\textit{ff}GT models of Ref.~\onlinecite{8}. The number $n$
of nuclides with experimental halflives below the prescribed limit is
given in the second column.  The quality indices labeling columns 3-8 are
defined in Eqs.~(\ref{eq:45}) -~(\ref{eq:50}).}
\begin{ruledtabular}
\begin{tabular}{llllllll}
\multicolumn{1}{c}{{ T$_{\beta,{\rm exp}}$}} & \multicolumn{7}{c}{{ (a) ANN Model. Overall Mode.}} \\ 
\multicolumn{1}{c}{${(s)}$} & \multicolumn{1}{c}{$n$} & \multicolumn{1}{c}{$M$} & \multicolumn{1}{c}{$M^{(10)}$} & \multicolumn{1}{c}{$\sigma_{\rm M}$} & \multicolumn{1}{c}{$\sigma_{\rm M^{(10)}}$} & \multicolumn{1}{c}{$\Sigma$} & \multicolumn{1}{c}{$\Sigma^{(10)}$} \\ 
\hline
\multicolumn{1}{l}{$<1$} & \multicolumn{1}{c}{252} & \multicolumn{1}{c}{0.09} & \multicolumn{1}{c}{1.24} & \multicolumn{1}{c}{0.39} & \multicolumn{1}{c}{2.44} & \multicolumn{1}{c}{0.40} & \multicolumn{1}{c}{2.50} \\ 
\multicolumn{1}{l}{$<10$} & \multicolumn{1}{c}{395} & \multicolumn{1}{c}{0.08} & \multicolumn{1}{c}{1.21} & \multicolumn{1}{c}{0.42} & \multicolumn{1}{c}{2.60} & \multicolumn{1}{c}{0.42} & \multicolumn{1}{c}{2.65} \\ 
\multicolumn{1}{l}{$<100$} & \multicolumn{1}{c}{529} & \multicolumn{1}{c}{0.07} & \multicolumn{1}{c}{1.17} & \multicolumn{1}{c}{0.43} & \multicolumn{1}{c}{2.68} & \multicolumn{1}{c}{0.43} & \multicolumn{1}{c}{2.71} \\ 
\multicolumn{1}{l}{$<1000$} & \multicolumn{1}{c}{653} & \multicolumn{1}{c}{0.07} & \multicolumn{1}{c}{1.18} & \multicolumn{1}{c}{0.45} & \multicolumn{1}{c}{2.84} & \multicolumn{1}{c}{0.46} & \multicolumn{1}{c}{2.88} \\ 
\multicolumn{1}{l}{$<10^6$} & \multicolumn{1}{c}{838} & \multicolumn{1}{c}{0.00} & \multicolumn{1}{c}{1.01} & \multicolumn{1}{c}{0.57} & \multicolumn{1}{c}{3.70} & \multicolumn{1}{c}{0.57} & \multicolumn{1}{c}{3.70} \\ 
\hline
\multicolumn{1}{c}{} & \multicolumn{1}{c}{} & \multicolumn{1}{c}{} & \multicolumn{1}{c}{} & \multicolumn{1}{c}{} & \multicolumn{1}{c}{} & \multicolumn{1}{c}{} & \multicolumn{1}{c}{} \\ 
\hline
\multicolumn{1}{c}{ $T_{\beta,{\rm exp}} $} & \multicolumn{7}{c}{ (b) ANN Model. Prediction Mode.} \\ 
\multicolumn{1}{c}{${(s)}$} & \multicolumn{1}{c}{$n$} & \multicolumn{1}{c}{$M$} & \multicolumn{1}{c}{$M^{(10)}$} & \multicolumn{1}{c}{$\sigma_{\rm M}$} & \multicolumn{1}{c}{$\sigma_{\rm M^{(10)}}$} & \multicolumn{1}{c}{$\Sigma$} & \multicolumn{1}{c}{$\Sigma^{(10)}$} \\ 
\hline
\multicolumn{1}{l}{$<1$} & \multicolumn{1}{c}{50} & \multicolumn{1}{c}{0.05} & \multicolumn{1}{c}{1.12} & \multicolumn{1}{c}{0.41} & \multicolumn{1}{c}{2.56} & \multicolumn{1}{c}{0.41} & \multicolumn{1}{c}{2.58} \\ 
\multicolumn{1}{l}{$<10$} & \multicolumn{1}{c}{79} & \multicolumn{1}{c}{0.02} & \multicolumn{1}{c}{1.05} & \multicolumn{1}{c}{0.48} & \multicolumn{1}{c}{3.00} & \multicolumn{1}{c}{0.48} & \multicolumn{1}{c}{3.01} \\ 
\multicolumn{1}{l}{$<100$} & \multicolumn{1}{c}{106} & \multicolumn{1}{c}{0.00} & \multicolumn{1}{c}{1.00} & \multicolumn{1}{c}{0.49} & \multicolumn{1}{c}{3.08} & \multicolumn{1}{c}{0.49} & \multicolumn{1}{c}{3.08} \\ 
\multicolumn{1}{l}{$<1000$} & \multicolumn{1}{c}{131} & \multicolumn{1}{c}{-0.03} & \multicolumn{1}{c}{0.93} & \multicolumn{1}{c}{0.50} & \multicolumn{1}{c}{3.16} & \multicolumn{1}{c}{0.50} & \multicolumn{1}{c}{3.17} \\ 
\multicolumn{1}{l}{$<10^6$} & \multicolumn{1}{c}{168} & \multicolumn{1}{c}{-0.09} & \multicolumn{1}{c}{0.82} & \multicolumn{1}{c}{0.64} & \multicolumn{1}{c}{4.38} & \multicolumn{1}{c}{0.65} & \multicolumn{1}{c}{4.44} \\ 
\hline
\multicolumn{1}{c}{} & \multicolumn{1}{c}{} & \multicolumn{1}{c}{} & \multicolumn{1}{c}{} & \multicolumn{1}{c}{} & \multicolumn{1}{c}{} & \multicolumn{1}{c}{} & \multicolumn{1}{c}{} \\ 
\hline
\multicolumn{1}{c}{{ T$_{\beta,{\rm exp}}$}} & \multicolumn{7}{c}{{ (c) $FRDM+$pn$QRPA$ Calculation~\cite{8}.}} \\ 
\multicolumn{1}{c}{${(s)}$} & \multicolumn{1}{c}{$n$} & \multicolumn{1}{c}{$M$} & \multicolumn{1}{c}{$M^{(10)}$} & \multicolumn{1}{c}{$\sigma_{\rm M}$} & \multicolumn{1}{c}{$\sigma_{\rm M^{(10)}}$} & \multicolumn{1}{c}{$\Sigma$} & \multicolumn{1}{c}{$\Sigma^{(10)}$} \\ 
\hline
\multicolumn{1}{l}{$<1$} & \multicolumn{1}{c}{184} & \multicolumn{1}{c}{0.03} & \multicolumn{1}{c}{1.06} & \multicolumn{1}{c}{0.57} & \multicolumn{1}{c}{3.72} & \multicolumn{1}{c}{0.57} & \multicolumn{1}{c}{3.73} \\ 
\multicolumn{1}{l}{$<10$} & \multicolumn{1}{c}{306} & \multicolumn{1}{c}{0.14} & \multicolumn{1}{c}{1.38} & \multicolumn{1}{c}{0.77} & \multicolumn{1}{c}{5.87} & \multicolumn{1}{c}{0.78} & \multicolumn{1}{c}{6.04} \\ 
\multicolumn{1}{l}{$<100$} & \multicolumn{1}{c}{431} & \multicolumn{1}{c}{0.19} & \multicolumn{1}{c}{1.55} & \multicolumn{1}{c}{0.94} & \multicolumn{1}{c}{8.81} & \multicolumn{1}{c}{0.96} & \multicolumn{1}{c}{9.21} \\ 
\multicolumn{1}{l}{$<1000$} & \multicolumn{1}{c}{546} & \multicolumn{1}{c}{0.34} & \multicolumn{1}{c}{2.20} & \multicolumn{1}{c}{1.28} & \multicolumn{1}{c}{19.09} & \multicolumn{1}{c}{1.33} & \multicolumn{1}{c}{21.17} \\ 
\multicolumn{1}{l}{$<10^6$} & \multicolumn{1}{c}{$-$} & \multicolumn{1}{c}{$-$} & \multicolumn{1}{c}{$-$} & \multicolumn{1}{c}{$-$} & \multicolumn{1}{c}{$-$} & \multicolumn{1}{c}{$-$} & \multicolumn{1}{c}{$-$} \\ 
\hline
\multicolumn{1}{c}{} & \multicolumn{1}{c}{} & \multicolumn{1}{c}{} & \multicolumn{1}{c}{} & \multicolumn{1}{c}{} & \multicolumn{1}{c}{} & \multicolumn{1}{c}{} & \multicolumn{1}{c}{} \\ 
\hline
\multicolumn{1}{c}{ \textbf{T$_{\beta,{\rm exp}}$}} & \multicolumn{7}{c}{ (d) $pn$QRPA +\textit{ff}GT Calculation~\cite{8}.} \\ 
\multicolumn{1}{c}{${(s)}$} & \multicolumn{1}{c}{$n$} & \multicolumn{1}{c}{$M$} & \multicolumn{1}{c}{$M^{(10)}$} & \multicolumn{1}{c}{$\sigma_{\rm M}$} & \multicolumn{1}{c}{$\sigma_{\rm M^{(10)}}$} & \multicolumn{1}{c}{$\Sigma$} & \multicolumn{1}{c}{$\Sigma^{(10)}$} \\ 
\hline
\multicolumn{1}{l}{$<1$} & \multicolumn{1}{c}{184} & \multicolumn{1}{c}{-0.08} & \multicolumn{1}{c}{0.84} & \multicolumn{1}{c}{0.48} & \multicolumn{1}{c}{3.04} & \multicolumn{1}{c}{0.49} & \multicolumn{1}{c}{3.08} \\ 
\multicolumn{1}{l}{$<10$} & \multicolumn{1}{c}{306} & \multicolumn{1}{c}{-0.03} & \multicolumn{1}{c}{0.93} & \multicolumn{1}{c}{0.55} & \multicolumn{1}{c}{3.52} & \multicolumn{1}{c}{0.55} & \multicolumn{1}{c}{3.53} \\ 
\multicolumn{1}{l}{$<100$} & \multicolumn{1}{c}{431} & \multicolumn{1}{c}{-0.04} & \multicolumn{1}{c}{0.91} & \multicolumn{1}{c}{0.61} & \multicolumn{1}{c}{4.10} & \multicolumn{1}{c}{0.61} & \multicolumn{1}{c}{4.12} \\ 
\multicolumn{1}{l}{$<1000$} & \multicolumn{1}{c}{546} & \multicolumn{1}{c}{-0.04} & \multicolumn{1}{c}{0.92} & \multicolumn{1}{c}{0.68} & \multicolumn{1}{c}{4.81} & \multicolumn{1}{c}{0.68} & \multicolumn{1}{c}{4.82} \\ 
\multicolumn{1}{l}{$<10^6$} & \multicolumn{1}{c}{$-$} & \multicolumn{1}{c}{$-$} & \multicolumn{1}{c}{$-$} & \multicolumn{1}{c}{$-$} & \multicolumn{1}{c}{$-$} & \multicolumn{1}{c}{$-$} & \multicolumn{1}{c}{$-$} \\ 
\end{tabular}
\end{ruledtabular}
\end{table}

\begin{table}
\caption{\label{tab:klap}
 Comparison of performance measures characterizing the  ANN model of the present
work, when operating in the Overall (a) and Prediction (b) Modes, with corresponding values for (c) the $pn$QRPA model of Staudt et al.~\cite{5} and (d) the NBCS+$pn$QRPA model of Homma et al.~\cite{6}. The quality indices $m\%$,  ${\bar x}_K$, and $\sigma_K$ are defined by Eqs.~(\ref{eq:41}) -~(\ref{eq:43}).  The third column reports the percentage $m\%$ of nuclides having experimental halflives within the prescribed range (second column), for which the calculated halflife lies within a certain tolerance factor (first column) of the experimental value.}
\begin{ruledtabular}
\begin{tabular}{ccccc}
 \multicolumn{5}{c}{ (a) ANN Model: Overall Mode.}\\
factor & $T_{\beta,{\rm exp}}$ (s) & $m\%$ & ${\bar x}_K$ & $\sigma_K$\\
\hline
\multicolumn{1}{l}{$<10$} & \multicolumn{1}{l}{$<10^6$} & \multicolumn{1}{c}{92.0} & \multicolumn{1}{c}{2.46} & \multicolumn{1}{c}{1.72} \\ 
\multicolumn{1}{c}{} & \multicolumn{1}{l}{$<60$} & \multicolumn{1}{c}{96.5} & \multicolumn{1}{c}{2.21} & \multicolumn{1}{c}{1.52} \\ 
\multicolumn{1}{c}{} & \multicolumn{1}{l}{$<1$} & \multicolumn{1}{c}{97.6} & \multicolumn{1}{c}{2.10} & \multicolumn{1}{c}{1.39} \\ 
\multicolumn{1}{l}{$<5$} & \multicolumn{1}{l}{$<10^6$} & \multicolumn{1}{c}{82.8} & \multicolumn{1}{c}{1.99} & \multicolumn{1}{c}{0.95} \\ 
\multicolumn{1}{c}{} & \multicolumn{1}{l}{$<60$} & \multicolumn{1}{c}{90.2} & \multicolumn{1}{c}{1.88} & \multicolumn{1}{c}{0.84} \\ 
\multicolumn{1}{c}{} & \multicolumn{1}{l}{$<1$} & \multicolumn{1}{c}{93.7} & \multicolumn{1}{c}{1.88} & \multicolumn{1}{c}{0.80} \\ 
\multicolumn{1}{l}{$<2$} & \multicolumn{1}{l}{$<10^6$} & \multicolumn{1}{c}{53.5} & \multicolumn{1}{c}{1.41} & \multicolumn{1}{c}{0.27} \\ 
\multicolumn{1}{c}{} & \multicolumn{1}{l}{$<60$} & \multicolumn{1}{c}{60.6} & \multicolumn{1}{c}{1.41} & \multicolumn{1}{c}{0.27} \\ 
\multicolumn{1}{c}{} & \multicolumn{1}{l}{$<1$} & \multicolumn{1}{c}{61.9} & \multicolumn{1}{c}{1.41} & \multicolumn{1}{c}{0.26} \\ 
\hline
\multicolumn{1}{c}{} & \multicolumn{1}{c}{} & \multicolumn{1}{c}{} & \multicolumn{1}{c}{} & \multicolumn{1}{c}{} \\ 
\hline
 \multicolumn{5}{c}{ (b) ANN Model:  Prediction Mode.}\\
factor & $T_{\beta,{\rm exp}}$ (s) & $m\%$ & ${\bar x}_K$ & $\sigma_K$\\
\hline
\multicolumn{1}{l}{$<10$} & \multicolumn{1}{l}{$<10^6$} & \multicolumn{1}{c}{90.5} & \multicolumn{1}{c}{2.69} & \multicolumn{1}{c}{1.85} \\ 
\multicolumn{1}{c}{} & \multicolumn{1}{l}{$<60$} & \multicolumn{1}{c}{96.1} & \multicolumn{1}{c}{2.48} & \multicolumn{1}{c}{1.64} \\ 
\multicolumn{1}{c}{} & \multicolumn{1}{l}{$<1$} & \multicolumn{1}{c}{98.0} & \multicolumn{1}{c}{2.24} & \multicolumn{1}{c}{1.30} \\ 
\multicolumn{1}{l}{$<5$} & \multicolumn{1}{l}{$<10^6$} & \multicolumn{1}{c}{79.2} & \multicolumn{1}{c}{2.10} & \multicolumn{1}{c}{0.97} \\ 
\multicolumn{1}{c}{} & \multicolumn{1}{l}{$<60$} & \multicolumn{1}{c}{87.3} & \multicolumn{1}{c}{2.05} & \multicolumn{1}{c}{0.91} \\ 
\multicolumn{1}{c}{} & \multicolumn{1}{l}{$<1$} & \multicolumn{1}{c}{94.0} & \multicolumn{1}{c}{2.04} & \multicolumn{1}{c}{0.89} \\ 
\multicolumn{1}{l}{$<2$} & \multicolumn{1}{l}{$<10^6$} & \multicolumn{1}{c}{49.4} & \multicolumn{1}{c}{1.48} & \multicolumn{1}{c}{0.28} \\ 
\multicolumn{1}{c}{} & \multicolumn{1}{l}{$<60$} & \multicolumn{1}{c}{53.9} & \multicolumn{1}{c}{1.48} & \multicolumn{1}{c}{0.27} \\ 
\multicolumn{1}{c}{} & \multicolumn{1}{l}{$<1$} & \multicolumn{1}{c}{60.0} & \multicolumn{1}{c}{1.50} & \multicolumn{1}{c}{0.27} \\ 
\hline
\multicolumn{1}{c}{} & \multicolumn{1}{c}{} & \multicolumn{1}{c}{} & \multicolumn{1}{c}{} & \multicolumn{1}{c}{} \\ 
\hline
\multicolumn{5}{c}{ (c) $pn$QRPA Calculation~\cite{5}.}\\
factor & $T_{\beta,{\rm exp}}$ (s) & $m\%$ & ${\bar x}_K$ & $\sigma_K$\\
\hline
\multicolumn{1}{l}{$<10$} & \multicolumn{1}{l}{$<10^6$} & \multicolumn{1}{c}{72.2} & \multicolumn{1}{c}{1.85} & \multicolumn{1}{c}{1.21} \\ 
\multicolumn{1}{c}{} & \multicolumn{1}{l}{$<60$} & \multicolumn{1}{c}{96.3} & \multicolumn{1}{c}{1.67} & \multicolumn{1}{c}{1.02} \\ 
\multicolumn{1}{c}{} & \multicolumn{1}{l}{$<1$} & \multicolumn{1}{c}{99.1} & \multicolumn{1}{c}{1.44} & \multicolumn{1}{c}{0.40} \\ 
\multicolumn{1}{l}{$<5$} & \multicolumn{1}{l}{$<10^6$} & \multicolumn{1}{c}{69.7} & \multicolumn{1}{c}{1.68} & \multicolumn{1}{c}{0.76} \\ 
\multicolumn{1}{c}{} & \multicolumn{1}{l}{$<60$} & \multicolumn{1}{c}{94.5} & \multicolumn{1}{c}{1.56} & \multicolumn{1}{c}{0.66} \\ 
\multicolumn{1}{c}{} & \multicolumn{1}{l}{$<1$} & \multicolumn{1}{c}{99.1} & \multicolumn{1}{c}{1.44} & \multicolumn{1}{c}{0.40} \\ 
\multicolumn{1}{l}{$<2$} & \multicolumn{1}{l}{$<10^6$} & \multicolumn{1}{c}{56.4} & \multicolumn{1}{c}{1.37} & \multicolumn{1}{c}{0.29} \\ 
\multicolumn{1}{c}{} & \multicolumn{1}{l}{$<60$} & \multicolumn{1}{c}{82.2} & \multicolumn{1}{c}{1.36} & \multicolumn{1}{c}{0.29} \\ 
\multicolumn{1}{c}{} & \multicolumn{1}{l}{$<1$} & \multicolumn{1}{c}{90.6} & \multicolumn{1}{c}{1.35} & \multicolumn{1}{c}{0.27} \\ 
\hline
\multicolumn{1}{c}{} & \multicolumn{1}{c}{} & \multicolumn{1}{c}{} & \multicolumn{1}{c}{} & \multicolumn{1}{c}{} \\ 
\hline
\multicolumn{5}{c}{ (d) NBCS+$pn$QRPA Calculation~\cite{6}.}\\
factor & $T_{\beta,{\rm exp}}$ (s) & $m\%$ & ${\bar x}_K$ & $\sigma_K$ \footnote{$\sigma_K$ results are not available in Ref.~\onlinecite{6}.}\\
\hline
\multicolumn{1}{l}{$<10$} & \multicolumn{1}{l}{$<10^6$} & \multicolumn{1}{c}{76.7} & \multicolumn{1}{c}{3.00} & \multicolumn{1}{c}{-} \\ 
\multicolumn{1}{c}{} & \multicolumn{1}{l}{$<60$} & \multicolumn{1}{c}{87.2} & \multicolumn{1}{c}{2.81} & \multicolumn{1}{c}{-} \\ 
\multicolumn{1}{c}{} & \multicolumn{1}{l}{$<1$} & \multicolumn{1}{c}{95.7} & \multicolumn{1}{c}{2.64} & \multicolumn{1}{c}{-} \\ 
\multicolumn{1}{l}{$<5$} & \multicolumn{1}{l}{$<10^6$} & \multicolumn{1}{c}{-} & \multicolumn{1}{c}{-} & \multicolumn{1}{c}{-} \\ 
\multicolumn{1}{c}{} & \multicolumn{1}{l}{$<60$} & \multicolumn{1}{c}{-} & \multicolumn{1}{c}{-} & \multicolumn{1}{c}{-} \\ 
\multicolumn{1}{c}{} & \multicolumn{1}{l}{$<1$} & \multicolumn{1}{c}{-} & \multicolumn{1}{c}{-} & \multicolumn{1}{c}{-} \\  
\multicolumn{1}{l}{$<2$} & \multicolumn{1}{l}{$<10^6$} & \multicolumn{1}{c}{33.8} & \multicolumn{1}{c}{1.43} & \multicolumn{1}{c}{-} \\ 
\multicolumn{1}{c}{} & \multicolumn{1}{l}{$<60$} & \multicolumn{1}{c}{42.0} & \multicolumn{1}{c}{1.41} & \multicolumn{1}{c}{-} \\ 
\multicolumn{1}{c}{} & \multicolumn{1}{l}{$<1$} & \multicolumn{1}{c}{50.7} & \multicolumn{1}{c}{1.43} & \multicolumn{1}{c}{-} \\ 
\end{tabular}
\end{ruledtabular}
\end{table}

\subsection{\label{sec:leve33}Comparison with RPA and GT Global Models - A Detailed Analysis} 

In this subsection, the performance of the favored network model of $\beta^-$ lifetime
systematics is compared with that of prominent theory-thick global models.
 
Adopting the quality measures (\ref{eq:45})--(\ref{eq:50}) introduced by M\"oller and collaborators, we first compare the performance of our global ANN model 
$\left[{ 3-5-5-5-5-1\left|116\right.} \right]$ with the global microscopic models based on the proton-neutron quasiparticle random-phase approximation ($pn$QRPA), in particular, the
NBCS+$pn$QRPA model of Homma et al.~\cite{6} and the FRDM+$pn$QRPA model of M\"{o}ller et al.~\cite{7}.  The efficacy of the ANN model is also compared with that of the micro-statistical Semi-Gross Theory (SGT) as implemented by Nakata et al.~\cite{12}.  Table~\ref{tab:odd-mass-pres} lists the 
ANN values for $M^{(10)}$ and $\sigma_M^{(10)}$ specific to odd-odd, odd-$A$, and even-even nuclides.  Table~\ref{tab:odd-mass-others} collects the $M^{(10)}$ and $\sigma_M^{(10)}$ values for the three theory-thick models in the same format.   As seen in these tables, both $pn$QRPA and SGT models tend to overestimate the $\beta^-$ halflives of odd-odd nuclei, while the FRDM calculation tends to underestimate the shorter halflives for even-even and odd mass nuclei. The ANN model, on the other hand, tends to overestimate the halflives of even-even nuclides,
although to a smaller degree; this shortcoming is due, at least in part, to the 
relative scarcity of even-even parents.

Table~\ref{tab:moller} contains values of the performance measures defined in
Eqs.~(\ref{eq:45})--(\ref{eq:50}) for three global models of $\beta^-$-decay
halflive.  Here the entries are not separated according to even-even, odd-$A$, or odd-odd class 
membership of the nuclides involved.  Included are results
for calculations within the FRDM+$pn$QRPA model, updated to a more recent
mass evaluation~\cite{8}, together with 
corresponding values for a hybrid ``micro-macroscopic'' $pn$QRPA+\textit{ff}GT treatment, which combines the QRPA model of allowed Gamow-Teller $\beta$ decay with the Gross Theory of first-forbidden ({\it ff}) decay~\cite{8}. In order to permit a direct comparison with the ANN model, we also report in this table the results for ANN performance figures determined independently of  the even-even, odd-$A$, odd-odd nuclidic class distinction, focusing attention only on the subdivision into halflife ranges.  The improved FRDM+$pn$QRPA model underestimates long halflives, whereas the $pn$QRPA+\textit{ff}GT approach slightly underestimates halflives over the full range considered.  
The tabulated quality indices indicate that the ANN responses are in closer agreement with experiment more frequently than the FRDM+$pn$QRPA  calculations, while the ANN model and the $pn$QRPA+\textit{ff}GT approaches perform about equally well.

The performance of our ANN model may also be evaluated in terms of the quality measures ${\bar x}_K$ and $\sigma_K$ employed by Klapdor and coworkers and defined in Eqs.~(\ref{eq:41})--(\ref{eq:43}). Table~\ref{tab:klap} includes values of these quantities for the network model,
along with values for the $pn$QRPA calculation of Staudt et al.~\cite{5} and for the NBCS+$pn$QRPA approach of Homma et al.~\cite{6}.  Detailed comparison shows that, judging from these indices, there
is only a modest decline in the quality of ANN responses in going from the Overall Mode to the Prediction Mode, and that the performance of the $pn$QRPA model is distinctly better than that of the neural network for shorter halflives but worse for longer halflife values.  We note, however,
that the $pn$QRPA model could be regarded as over-parameterized compared to more up-to-date models, since the strengths of the {\it NN} interactions are derived from a local fitting of the experimental data in each  chain. Turning to the NBCS+$pn$QRPA calculation, 
it is evident from Table~\ref{tab:klap} that the ANN model generally exhibits smaller discrepancies between calculated and observed $\beta^-$-decay halflives.  For example, the network model has the ability to reproduce approximately 50\% of experimentally known halflives shorter than $10^6$ $s$ within a factor of 2.  It should be noted, however, that the NBCS+$pn$QRPA model has fewer adjustable parameters~\cite{6}.  
     
Viewed as a whole, the analyses presented in Tables \ref{tab:odd-mass-pres}-\ref{tab:klap} demonstrate that in a clear majority of cases in which the statistical model of $\beta^-$ halflives is presented with test nuclides {\it absent} from the training and validation sets, it
makes {\it predictions} that are closer to experiment than the corresponding results from traditional models based on quantum many-body theory and phenomenology. This is ascribed to some extend to the larger number of adjustable parameters of the current model.

\begin{table}
\caption{\label{tab:dakos}
Performance measures for the $\left[{16-10-1\left|181\right.} \right]$ ANN model
constructed by Mavrommatis et al.~\cite{13}.  The quality indices ${\bar x}_K$ and
$\sigma_K$, introduced by Klapdor and coworkers, are defined in
Eqs.~(\ref{eq:41}) and~(\ref{eq:43}), respectively, while $m\%$ is the percentage of nuclides having experimental halflives within the prescribed range (second column), for which the calculated halflife lies within the tolerance factor (first column) 
of the experimental value.}
\begin{ruledtabular}
\begin{tabular}{lllll}
\multicolumn{5}{c}{ Prediction Mode. \rm{ANN} model of Ref.~\onlinecite{13}.}  \\ 
\multicolumn{1}{c}{factor} & \multicolumn{1}{c}{$T_{\beta,{\rm exp}}$ (s)} & \multicolumn{1}{c}{$m\%$} & \multicolumn{1}{c}{${\bar x}_K$} & \multicolumn{1}{c}{$\sigma_K$} \\ 
\hline
\multicolumn{1}{l}{$<10$} & \multicolumn{1}{l}{$<10^6$} & \multicolumn{1}{c}{82.8} & \multicolumn{1}{c}{2.78} & \multicolumn{1}{c}{1.83} \\ 
\multicolumn{1}{c}{ } & \multicolumn{1}{l}{$<60$} & \multicolumn{1}{c}{88.1} & \multicolumn{1}{c}{2.80} & \multicolumn{1}{c}{1.83} \\ 
\multicolumn{1}{c}{ } & \multicolumn{1}{l}{$<1$} & \multicolumn{1}{c}{90.0} & \multicolumn{1}{c}{2.88} & \multicolumn{1}{c}{1.88} \\ 
\multicolumn{1}{l}{$<5$} & \multicolumn{1}{l}{$<10^6$} & \multicolumn{1}{c}{72.4} & \multicolumn{1}{c}{2.22} & \multicolumn{1}{c}{1.07} \\ 
\multicolumn{1}{c}{ } & \multicolumn{1}{l}{$<60$} & \multicolumn{1}{c}{76.2} & \multicolumn{1}{c}{2.20} & \multicolumn{1}{c}{1.01} \\ 
\multicolumn{1}{c}{ } & \multicolumn{1}{l}{$<1$} & \multicolumn{1}{c}{76.7} & \multicolumn{1}{c}{2.23} & \multicolumn{1}{c}{1.02} \\ 
\multicolumn{1}{l}{$<2$} & \multicolumn{1}{l}{$<10^6$} & \multicolumn{1}{c}{39.7} & \multicolumn{1}{c}{1.39} & \multicolumn{1}{c}{0.29} \\ 
\multicolumn{1}{c}{ } & \multicolumn{1}{l}{$<60$} & \multicolumn{1}{c}{42.9} & \multicolumn{1}{c}{1.44} & \multicolumn{1}{c}{0.32} \\ 
\multicolumn{1}{c}{ } & \multicolumn{1}{l}{$<1$} & \multicolumn{1}{c}{43.3} & \multicolumn{1}{c}{1.46} & \multicolumn{1}{c}{0.32} \\ 
\end{tabular}
\end{ruledtabular}
\end{table}

\begin{table}
\caption{\label{tab:clark} Performance measures for the $\left[{17-10-1\left|191\right.} \right]$ ANN model constructed by Clark et al.~\cite{14}.  The quality indices $M^{(10)}$ and $\sigma_{\rm M^{(10)}}$, introduced by M\"oller and coworkers, are defined in Eqs.~(\ref{eq:48})--(\ref{eq:49}).}
\begin{ruledtabular}
\begin{tabular}{llll}
\multicolumn{1}{c}{$ T_{\beta,{\rm exp}}$} & \multicolumn{3}{c}{ Prediction Mode. \rm{ANN} model of Ref.~\onlinecite{14}.} \\ 
\multicolumn{1}{c}{(s)} & \multicolumn{1}{c}{Class} & \multicolumn{1}{c}{$M^{(10)}$} & \multicolumn{1}{c}{$\sigma_{\rm M^{(10)}}$} \\ 
\hline
\multicolumn{1}{l}{$<1$} & \multicolumn{1}{c}{o-o} & \multicolumn{1}{c}{2.05} & \multicolumn{1}{c}{2.31} \\ 
\multicolumn{1}{c}{ } & \multicolumn{1}{c}{odd} & \multicolumn{1}{c}{1.08} & \multicolumn{1}{c}{2.38} \\ 
\multicolumn{1}{c}{ } &\multicolumn{1}{c}{e-e} & \multicolumn{1}{c}{1.79} & \multicolumn{1}{c}{2.71} \\ 
\multicolumn{1}{c}{ } & \multicolumn{1}{c}{} & \multicolumn{1}{c}{} & \multicolumn{1}{c}{} \\ 
\multicolumn{1}{l}{$<10$} & \multicolumn{1}{c}{o-o} & \multicolumn{1}{c}{2.26} & \multicolumn{1}{c}{5.42} \\ 
\multicolumn{1}{c}{ } & \multicolumn{1}{c}{odd} & \multicolumn{1}{c}{1.19} & \multicolumn{1}{c}{2.44} \\ 
\multicolumn{1}{c}{ } & \multicolumn{1}{c}{e-e} & \multicolumn{1}{c}{1.31} & \multicolumn{1}{c}{2.30} \\ 
\multicolumn{1}{c}{ } & \multicolumn{1}{c}{} & \multicolumn{1}{c}{} & \multicolumn{1}{c}{} \\ 
\multicolumn{1}{l}{$<100$} & \multicolumn{1}{c}{o-o} & \multicolumn{1}{c}{1.76} & \multicolumn{1}{c}{5.19} \\ 
\multicolumn{1}{c}{ } & \multicolumn{1}{c}{odd} & \multicolumn{1}{c}{1.12} & \multicolumn{1}{c}{3.15} \\ 
\multicolumn{1}{c}{ } & \multicolumn{1}{c}{e-e}  & \multicolumn{1}{c}{0.98} & \multicolumn{1}{c}{2.67} \\ 
\multicolumn{1}{c}{ } & \multicolumn{1}{c}{} & \multicolumn{1}{c}{} & \multicolumn{1}{c}{} \\ 
\multicolumn{1}{l}{$<1000$} & \multicolumn{1}{c}{o-o} & \multicolumn{1}{c}{2.22} & \multicolumn{1}{c}{6.25} \\ 
\multicolumn{1}{c}{ } & \multicolumn{1}{c}{odd} & \multicolumn{1}{c}{1.22} & \multicolumn{1}{c}{5.50} \\ 
\multicolumn{1}{c}{ } & \multicolumn{1}{c}{e-e} & \multicolumn{1}{c}{0.93} & \multicolumn{1}{c}{4.78} \\ 
\end{tabular}
\end{ruledtabular}
\end{table}

\begin{table}
\caption{\label{tab:li}  Root-mean-square errors ($\sigma_{\rm RMSE}$) for 
(a) the $\left[{3-5-5-5-5-1\left|116\right.} \right]$ ANN  model of the present work, and  (b)  the SVM model constructed by Li et al.~\cite{25}.  Here $n$ is the number of nuclides  in each of the data (sub)sets.
}
 \begin{ruledtabular}
\begin{tabular}{lllllll}
\multicolumn{7}{c}{ (a) ANN Model.} \\ 
\hline
\multicolumn{1}{c}{} & \multicolumn{2}{c}{Learning Set  } & \multicolumn{2}{c}{Validation Set } & \multicolumn{2}{c}{Test Set} \\ 
\multicolumn{1}{c}{Class} & \multicolumn{1}{c}{$n$} & \multicolumn{1}{c}{$\sigma_{\rm RMSE}$} & \multicolumn{1}{c}{$n$} & \multicolumn{1}{c}{$\sigma_{\rm RMSE}$} & \multicolumn{1}{c}{$n$} & \multicolumn{1}{c}{$\sigma_{\rm RMSE}$} \\ 
\hline
\multicolumn{1}{c}{EE} & \multicolumn{1}{c}{95} & \multicolumn{1}{c}{0.52} & \multicolumn{1}{c}{33} & \multicolumn{1}{c}{0.52} & \multicolumn{1}{c}{35} & \multicolumn{1}{c}{0.64} \\ 
\multicolumn{1}{c}{EO} & \multicolumn{1}{c}{121} & \multicolumn{1}{c}{0.55} & \multicolumn{1}{c}{46} & \multicolumn{1}{c}{0.77} & \multicolumn{1}{c}{47} & \multicolumn{1}{c}{0.57} \\ 
\multicolumn{1}{c}{OE} & \multicolumn{1}{c}{141} & \multicolumn{1}{c}{0.46} & \multicolumn{1}{c}{42} & \multicolumn{1}{c}{0.53} & \multicolumn{1}{c}{40} & \multicolumn{1}{c}{0.66} \\ 
\multicolumn{1}{c}{OO} & \multicolumn{1}{c}{146} & \multicolumn{1}{c}{0.56} & \multicolumn{1}{c}{46} & \multicolumn{1}{c}{0.52} & \multicolumn{1}{c}{46} & \multicolumn{1}{c}{0.71} \\ 
\multicolumn{1}{c}{{Total}} & \multicolumn{1}{c}{{503}} & \multicolumn{1}{c}{{0.53}} & \multicolumn{1}{c}{{167}} & \multicolumn{1}{c}{{0.58}} & \multicolumn{1}{c}{{168}} & \multicolumn{1}{c}{{0.65}} \\ 
\hline
\multicolumn{1}{c}{} & \multicolumn{1}{c}{} & \multicolumn{1}{c}{} & \multicolumn{1}{c}{} & \multicolumn{1}{c}{} & \multicolumn{1}{c}{} & \multicolumn{1}{c}{} \\ 
\multicolumn{7}{c}{ (b) SVMs Calculation. Li et al.~\cite{25}.} \\ 
\hline
\multicolumn{1}{c}{} & \multicolumn{2}{c}{Learning Set} & \multicolumn{2}{c}{Validation Set} & \multicolumn{2}{c}{Test Set} \\ 
\multicolumn{1}{c}{Class} & \multicolumn{1}{c}{$n$} & \multicolumn{1}{c}{$\sigma_{\rm RMSE}$} & \multicolumn{1}{c}{$n$} & \multicolumn{1}{c}{$\sigma_{\rm RMSE}$} & \multicolumn{1}{c}{$n$} & \multicolumn{1}{c}{$\sigma_{\rm RMSE}$} \\ 
\hline
\multicolumn{1}{c}{EE} & \multicolumn{1}{c}{131} & \multicolumn{1}{c}{0.55} & \multicolumn{1}{c}{16} & \multicolumn{1}{c}{0.57} & \multicolumn{1}{c}{16} & \multicolumn{1}{c}{0.62} \\ 
\multicolumn{1}{c}{EO} & \multicolumn{1}{c}{179} & \multicolumn{1}{c}{0.41} & \multicolumn{1}{c}{22} & \multicolumn{1}{c}{0.42} & \multicolumn{1}{c}{22} & \multicolumn{1}{c}{0.51} \\ 
\multicolumn{1}{c}{OE} & \multicolumn{1}{c}{172} & \multicolumn{1}{c}{0.41} & \multicolumn{1}{c}{21} & \multicolumn{1}{c}{0.47} & \multicolumn{1}{c}{21} & \multicolumn{1}{c}{0.47} \\ 
\multicolumn{1}{c}{OO} & \multicolumn{1}{c}{190} & \multicolumn{1}{c}{0.52} & \multicolumn{1}{c}{24} & \multicolumn{1}{c}{0.4} & \multicolumn{1}{c}{24} & \multicolumn{1}{c}{0.52} \\ 
\multicolumn{1}{c}{{Total}} & \multicolumn{1}{c}{{672}} & \multicolumn{1}{c}{{0.47}} & \multicolumn{1}{c}{{83}} & \multicolumn{1}{c}{{0.46}} & \multicolumn{1}{c}{{83}} & \multicolumn{1}{c}{{0.53}} \\ 
\end{tabular}
\end{ruledtabular}
\end{table}

\subsection{\label{sec:leve34}Comparison with Prior ANN and SVM Models}

Some exploratory applications of artificial neural networks
to $\beta$-decay systematics were carried out earlier by the 
Athens-Manchester-St.~Louis collaboration and reported
in Refs.~\onlinecite{13,14}.  The first of these studies arrived at a 
fully-connected multilayer
feedforward ANN model having the simple architecture $\left[{16-10-1\left|181\right.} \right]$, and the second dealt with a similar model with architecture
$\left[{17-10-1\left|191\right.} \right]$.
Both of these efforts employed binary encoding of $Z$ and $N$ at the input, used the same data sets which differed from the ones of the present work and
implemented a quite orthodox backpropagation algorithm, incorporating a 
momentum term to enhance convergence of the learning process \cite{15}.
The main difference between these two earlier ANN models is the addition,
in the second, of an analog input unit representing the $Q$-value
of the decay.  Tables~\ref{tab:dakos} and \ref{tab:clark} present values
for performance measures of these ANN models operating in the Prediction
Mode.  (We concentrate on this aspect of performance, since it relates directly to
the \textit{extrapability} of the models.)   For the  $\left[{16-10-1\left|181\right.} \right]$ network 
model, Table~\ref{tab:dakos} displays results for the quality measures used by Klapdor
and coworkers, evaluated on the test set.   For the $\left[{17-10-1\left|191\right.} \right]$
model, Table~\ref{tab:clark} gives results for the performance measures of M\"oller
and coworkers, based on the responses of the model to the same test set.
Upon comparison with the entries for $M^{(10)}$ in Table~\ref{tab:odd-mass-pres},
one sees that the performance of the 17-input network model is rather similar
to that of the present 6-layer ANN model, except for odd-odd nuclides -- whose
lifetimes are overestimated by the older network.  In the case of the 16-input
model, comparison of the entries for $m$\% in Tables~\ref{tab:dakos} and
\ref{tab:klap} provides substantial evidence for the superiority of 
the new ANN model developed here, although this is not so clearly reflected
in the respective ${\bar x}_K$ values.

From a strategic standpoint, the advantages of the current ANN model over the earlier ones are twofold.   First, the number of degrees of freedom (weight and bias parameters) is reduced considerably by the use
of analog encoding of $Z$ and $N$.  Despite the greater number
of hidden layers, the current model, with architecture
$\left[{3-5-5-5-5-1\left|116\right.} \right]$, 
has 65 parameters fewer than the 16-input model and 75 less than
the 17-input model.  Secondly, there is the advantage relative to the  
latter model that the current version does not rely on $Q$-value input.
Experimental $Q$-values are not known for all the nuclides of interest, 
so the need to call upon theoretical results for input variables is eliminated.

As mentioned in the introduction,
initial studies of the classification and regression problems presented by
nuclear systematics have recently been carried out \cite{25,55} using the relatively
new methodology of Support Vector Machines (SVMs).  SVMs, which belong
to the class of kernel methods \cite{15}, are
learning systems having a rigorous basis in the statistical learning theory developed
by Vapnick and Chervonenkis~\cite{27} (VC theory).  There are similarities to
multilayer feedforward neural networks, notably in architecture, but there are
also important differences having to do with the better control over the tradeoff
between complexity and generalization ability within the SVM framework.
Importantly, within this framework there is an automated process for 
determining the explicit weights of the network in terms of a set of support vectors
optimally distilled from among the training patterns \cite{26}.   The few remaining 
parameters are embodied in the inner-product kernel that allows one to deal
efficiently with the high-dimensional feature space appropriate to the problem
to be solved.  The SVM methodology was originally developed for classification
problems, but has been extended to function approximation (regression)
\cite{15}.

The recent applications of SVMs to global modeling of nuclear properties, including
atomic masses, $\alpha$ decay chains of superheavy nuclei, ground-state spins and parities, and $\beta^-$ lifetimes, demonstrate considerable promise for this
approach.  As in the present work, cross-validation is performed, separating
the full database into learning, validation, and test sets.  In the existing studies,
the data for a given property is divided into four nonoverlapping subsets
containing input-output pairs for even-even, even-odd, odd-even, and odd-odd
classes of nuclides distinguished by the parity of $Z$ and $N$.

Table~\ref{tab:li} provides values of the conventional $\sigma_{\rm RMSE}$ performance
measure (\ref{eq:29}), both for the SVM model of $\beta^-$-decay
systematics constructed by Clark et al.~\cite{25} and for the present
ANN model.   The SVM model demonstrates better performance based
on this comparison, with a few exceptions involving the even-even
nuclides.  However, this comparison is somewhat misleading, since a larger fraction of the data was used for 
training, leaving numerically smaller validation and test sets in the SVM construction.  It must be noted in this regard that the
subdivision of the nuclides into four $(Z,N)$ parity classes requires four
separate SVM approximation processes to be executed.  This can lead to
spurious fluctuations in the predictions of lifetimes for nuclides
of isotopic and isotonic chains, as found in detailed inspection of
the outputs of the SVM model.  We should note further, however, that
a subsequent SVM model of $\beta^-$ systematics shows $\sigma_{\rm RMSE}$
values significantly lower than those given in Table~\ref{tab:li} for
the SVM model of Li et al.

\begin{figure}

\includegraphics[width=3.45in]{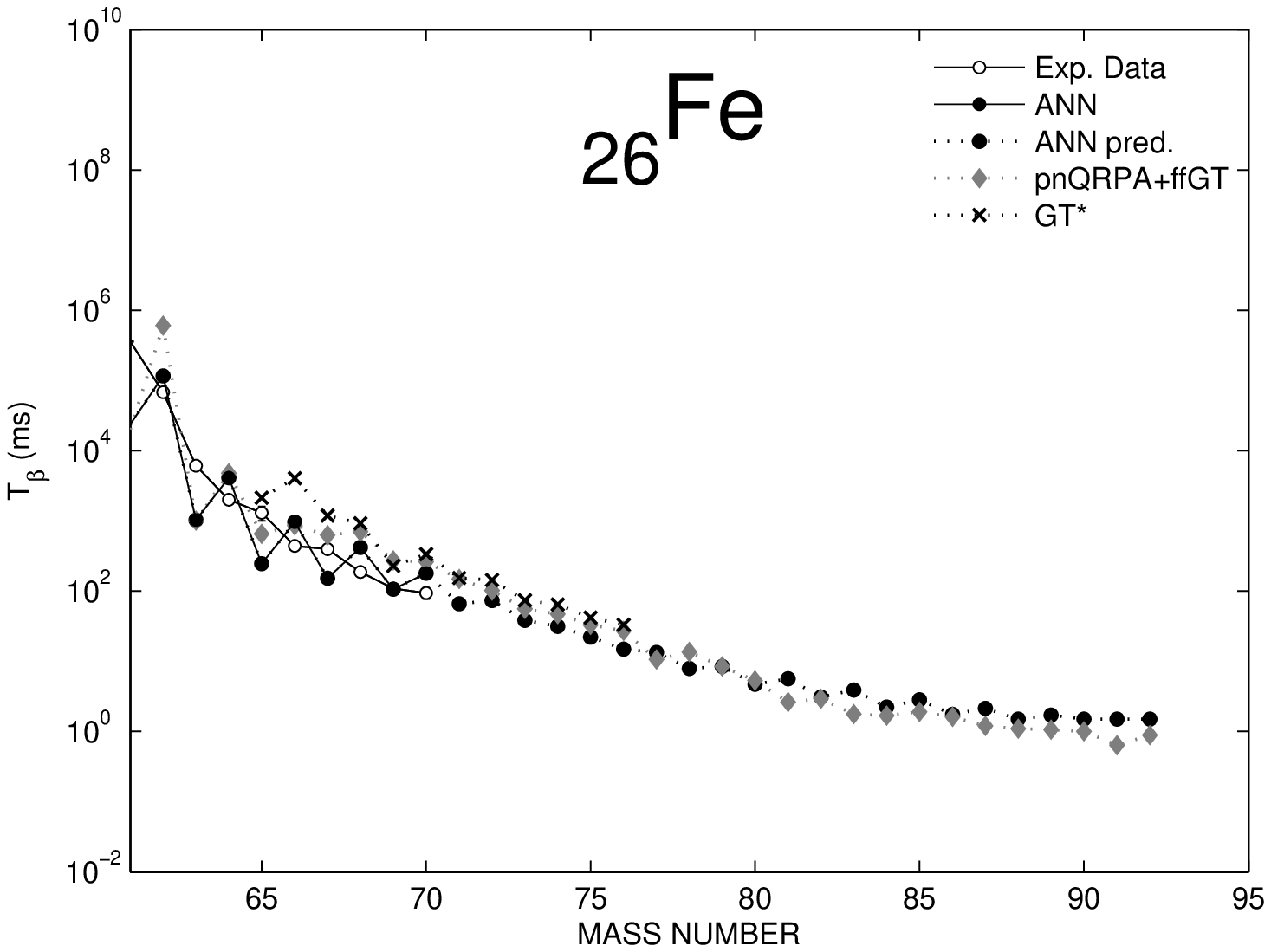}
\caption{\label{fig:fe26} Experimental data and derived halflives from different models for the isotopic chain of $_{26}$\rm{Fe}.}

\noindent{}

\includegraphics[width=3.45in]{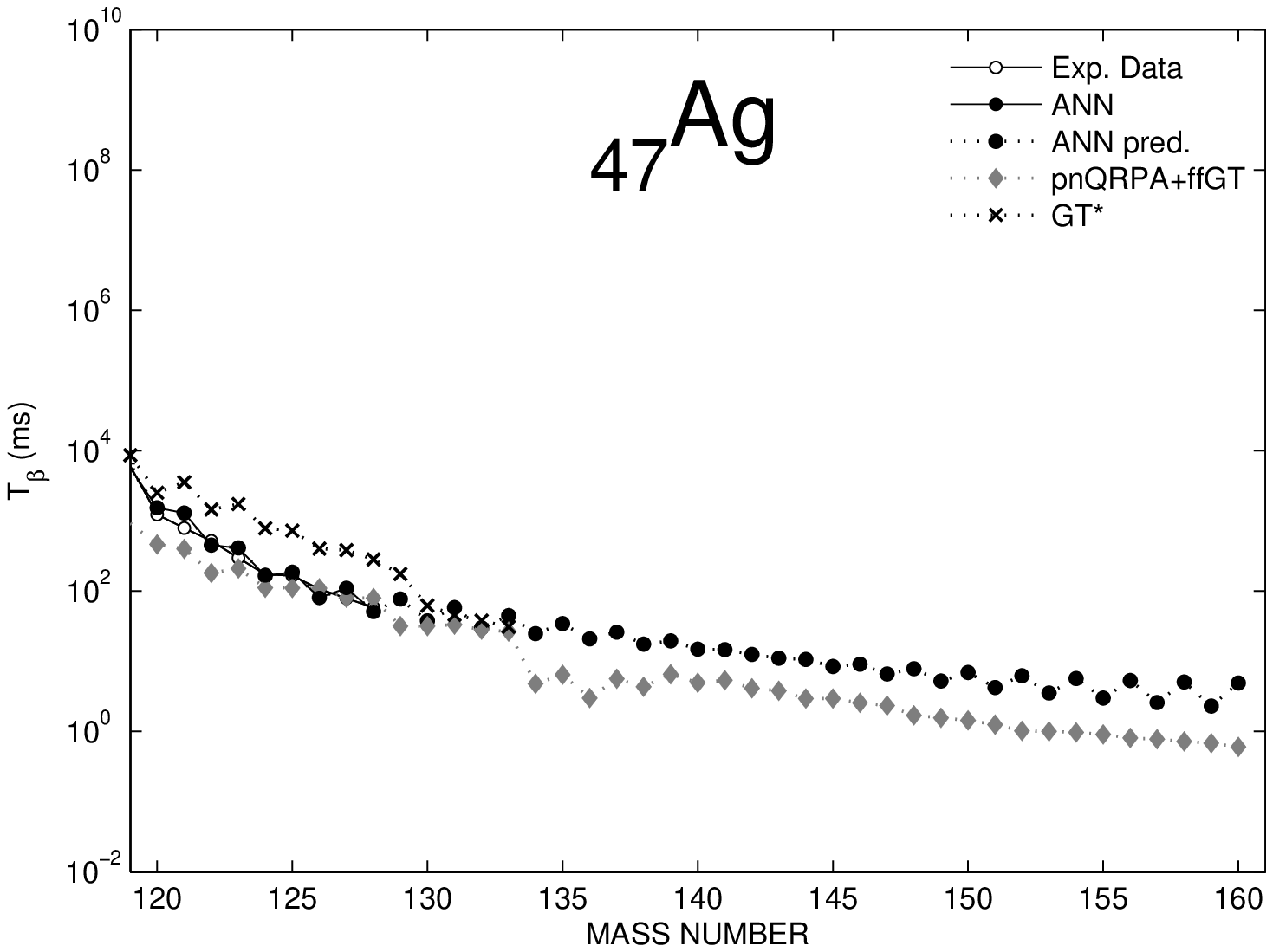}
\caption{\label{fig:ag47} The same as in Fig.~\ref{fig:fe26} but for the isotopic chain of $_{47}$\rm{Ag}.}

\noindent{}

\includegraphics[width=3.45in]{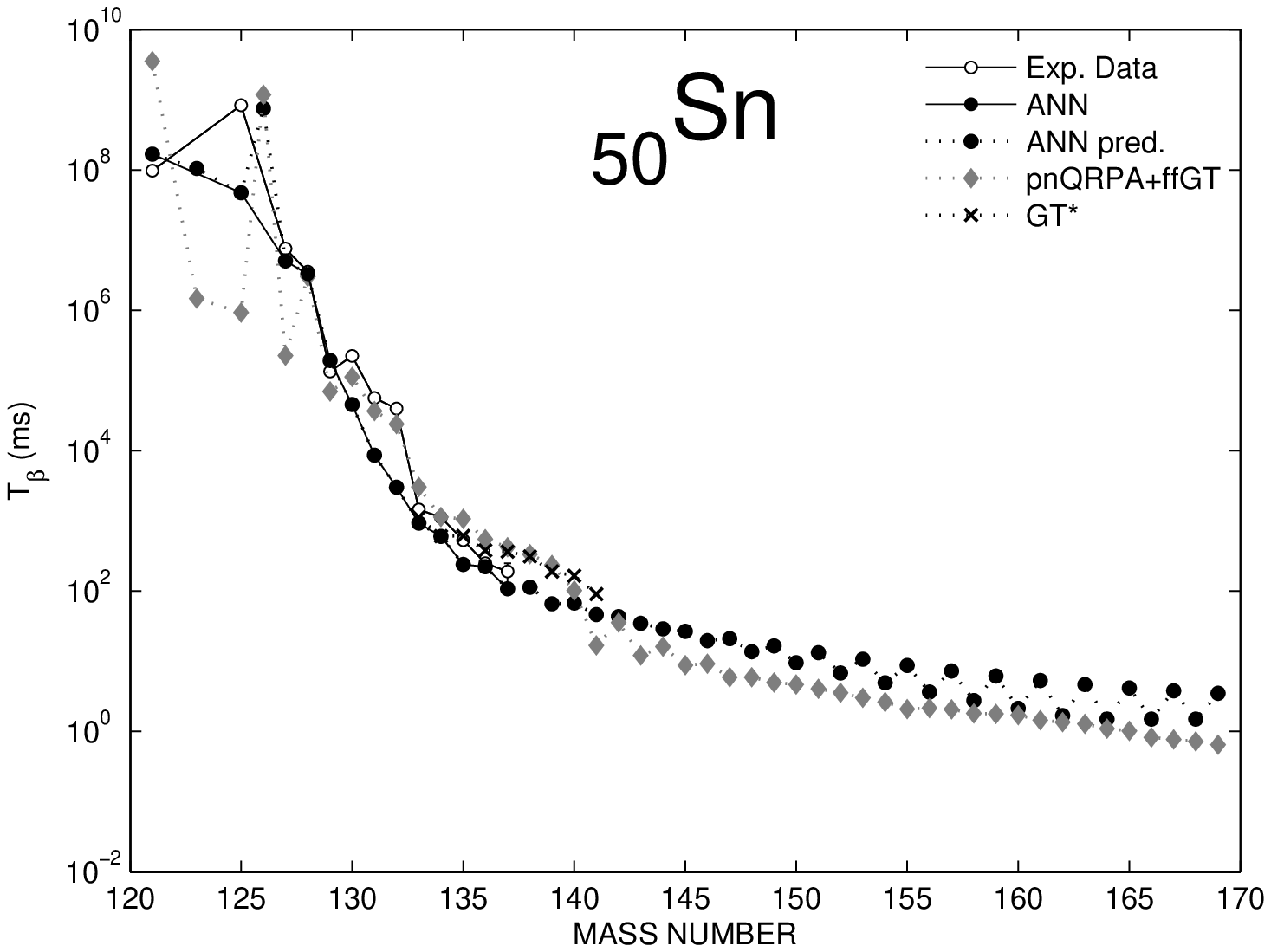}
\caption{\label{fig:sn50} The same as in Fig.~\ref{fig:fe26} but for the isotopic chain of $_{50}$\rm{Sn}.}

\end{figure}

\begin{figure}

\includegraphics[width=3.45in]{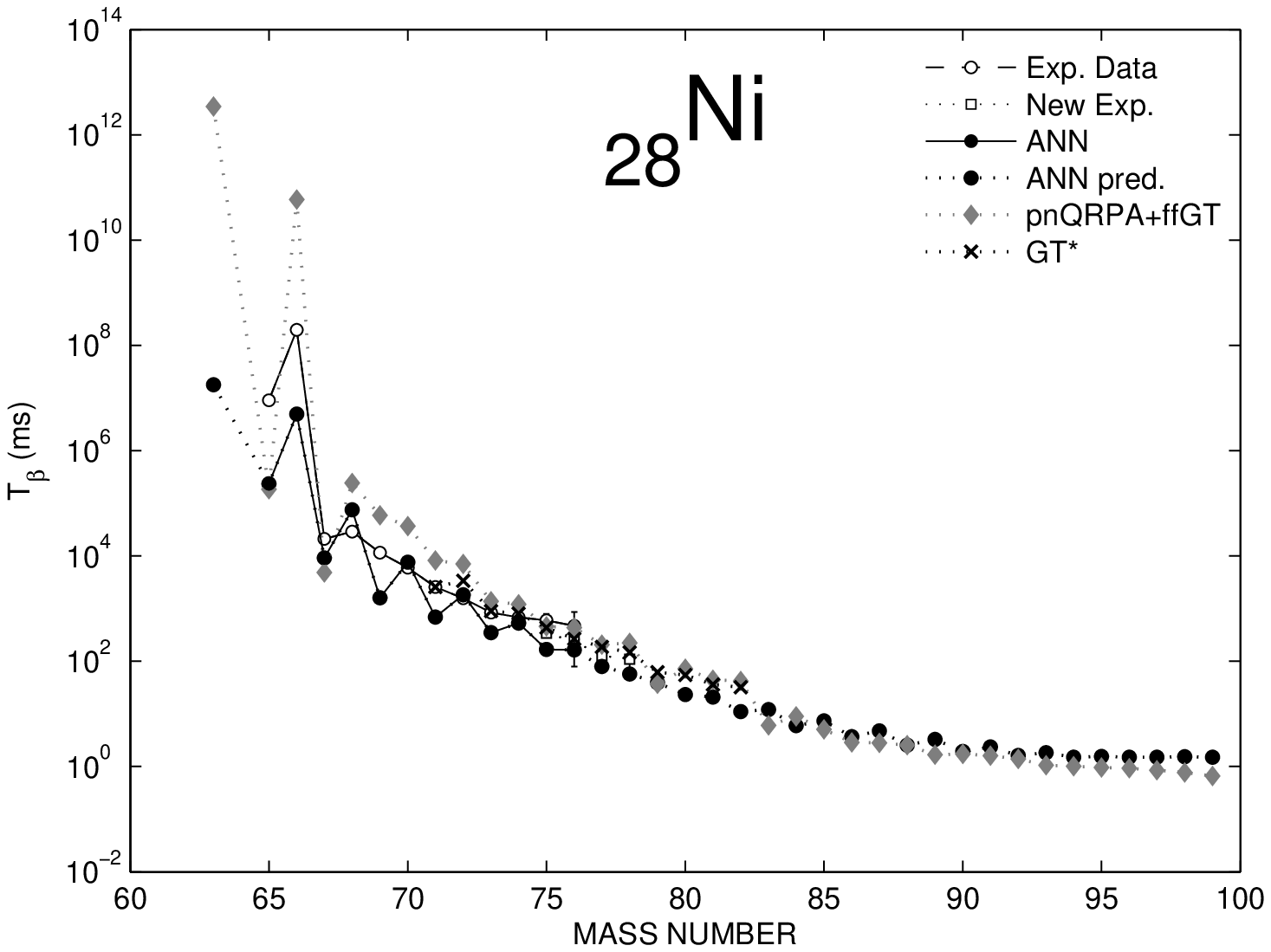}
\caption{\label{fig:ni28} The same as in Fig.~\ref{fig:fe26} but for the isotopic chain of
$_{28}$\rm{Ni}.}

\noindent{}

\includegraphics[width=3.45in]{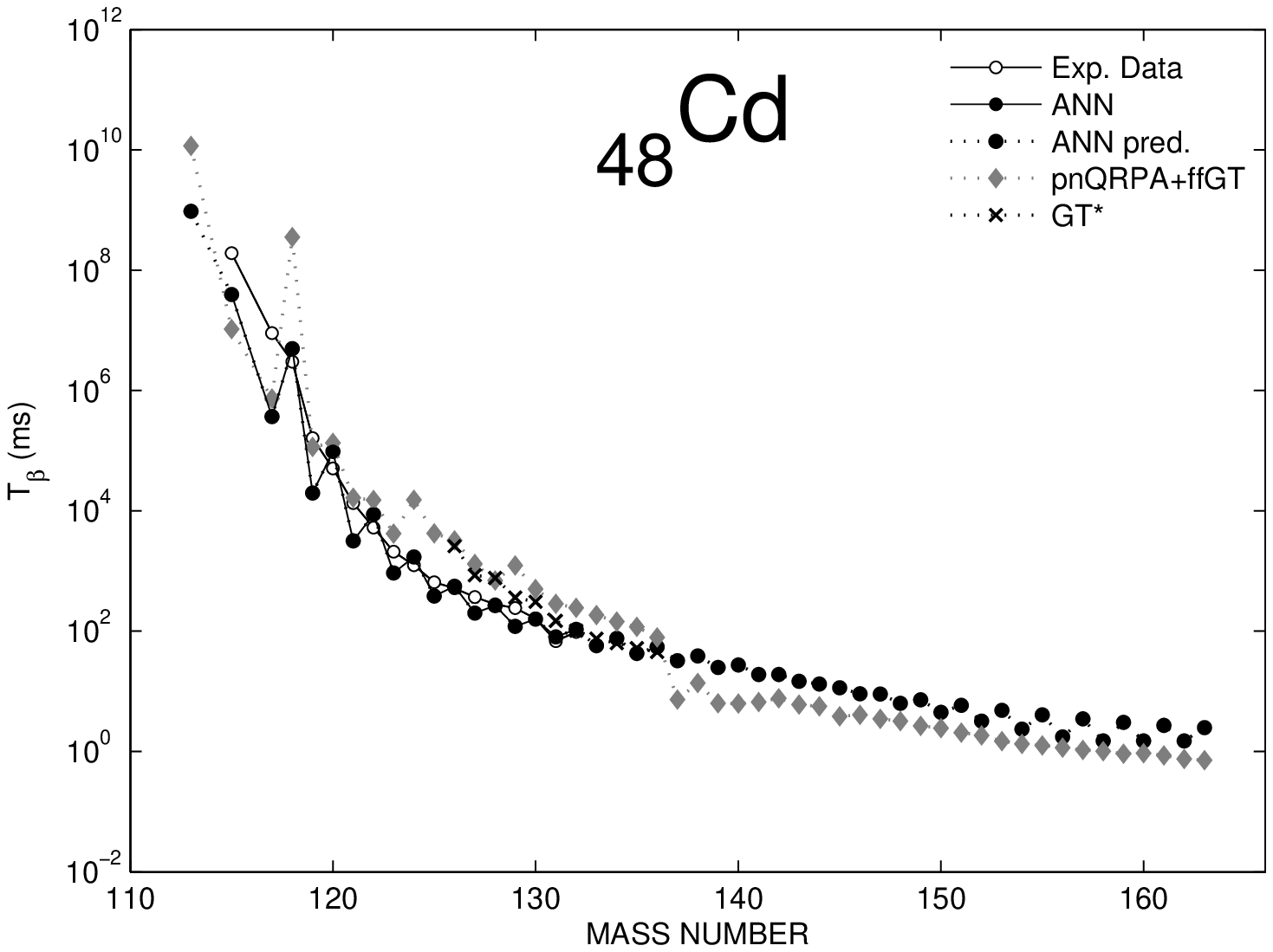}
\caption{\label{fig:cd48} The same as in Fig.~\ref{fig:fe26} but for the isotopic chain of $_{48}$\rm{Cd}.}

\noindent{}

\includegraphics[width=3.45in]{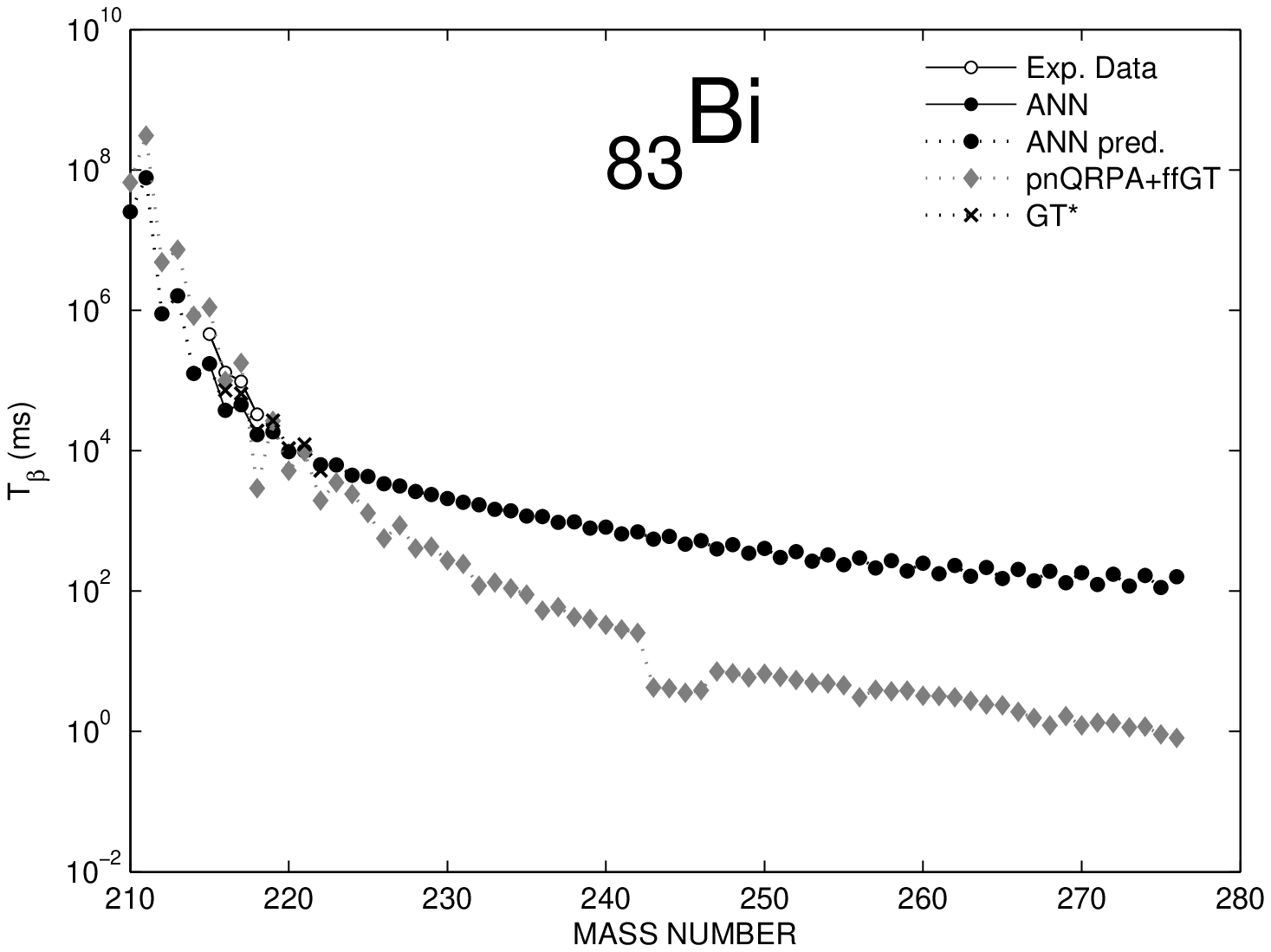}
\caption{\label{fig:bi83} The same as in Fig.~\ref{fig:fe26} but for the isotopic chain of 
$_{83}$\rm{Bi}.}

\end{figure}

\subsection{\label{sec:leve35} The Extrapability of the ANN Model}

It is of course desirable to have a model that reproduces experimentally 
known $\beta^-$ halflives of nuclei across the known nuclear landscape.
One can certainly achieve that goal with a sufficiently complex model
that involves a sufficient number of adjustable parameters.  However,
excess complexity generally implies poor predictive ability, and 
especially poor extrapability -- lack of the ability to extrapolate away
from existing data.  Accordingly, a much more important and challenging
goal is to develop a global model, statistical or otherwise, with minimal 
complexity consistent with good generalization properties.  The
extent to which this goal can be achieved with machine-learning
techniques for different nuclear properties is yet to be decided.  Of
course, one can test the performance of a favored network model on
outlying nuclei (outlying with respect to the valley of stability), 
nuclei that are unknown to the network, but have known 
values for the property of interest.  Adequate performance
in such tests can provide some degree of confidence in
predictions made by the model for nearby nuclei that have not
yet been reached by experiment.

In this subsection, we present some specific evidence of the
extrapability of the $\left[{3-5-5-5-5-1\left|116\right.} \right]$ ANN  
model developed in the present work. Figs.~10--15
show the halflives estimated by the model for nuclides in the \rm{Fe}, \rm{Ag}, \rm{Sn}, \rm{Ni}, \rm{Cd}, and \rm{Bi} isotopic chains.   Corresponding $pn$QRPA+\textit{ff}GT estimates are included for a comparison.  Also included are some results (labeled GT*) from calculations by Pfeiffer, Kratz, and M\"{o}ller~\cite{77} based on the early Gross Theory (GT) of Takahashi 
et al.~\cite{22}, with updated mass values~\cite{23,24} (GT*).  There is no unambiguous criterion that can be used to gauge the performance of these models.  Judging from the observed behavior of the known nuclei, one can generally expect that the more neutron-rich an exotic isotope is, the shorter its halflife.  This expected downward tendency is
predicted by all the models.  
One also expects to see some even-odd stagger of the points for
neighboring isotopes.
The ANN model produces such behavior, but it is probably overestimated.
Similar behavior, though less pronounced, appears in the results from continuum-Quasiparticle-RPA (CQRPA) approaches~\cite{56} and in the results of other theoretical calculations~\cite{8,22}.

\subsection{\label{sec:leve36} The r-Process Path}

\begin{figure}

\includegraphics[width=3.45in]{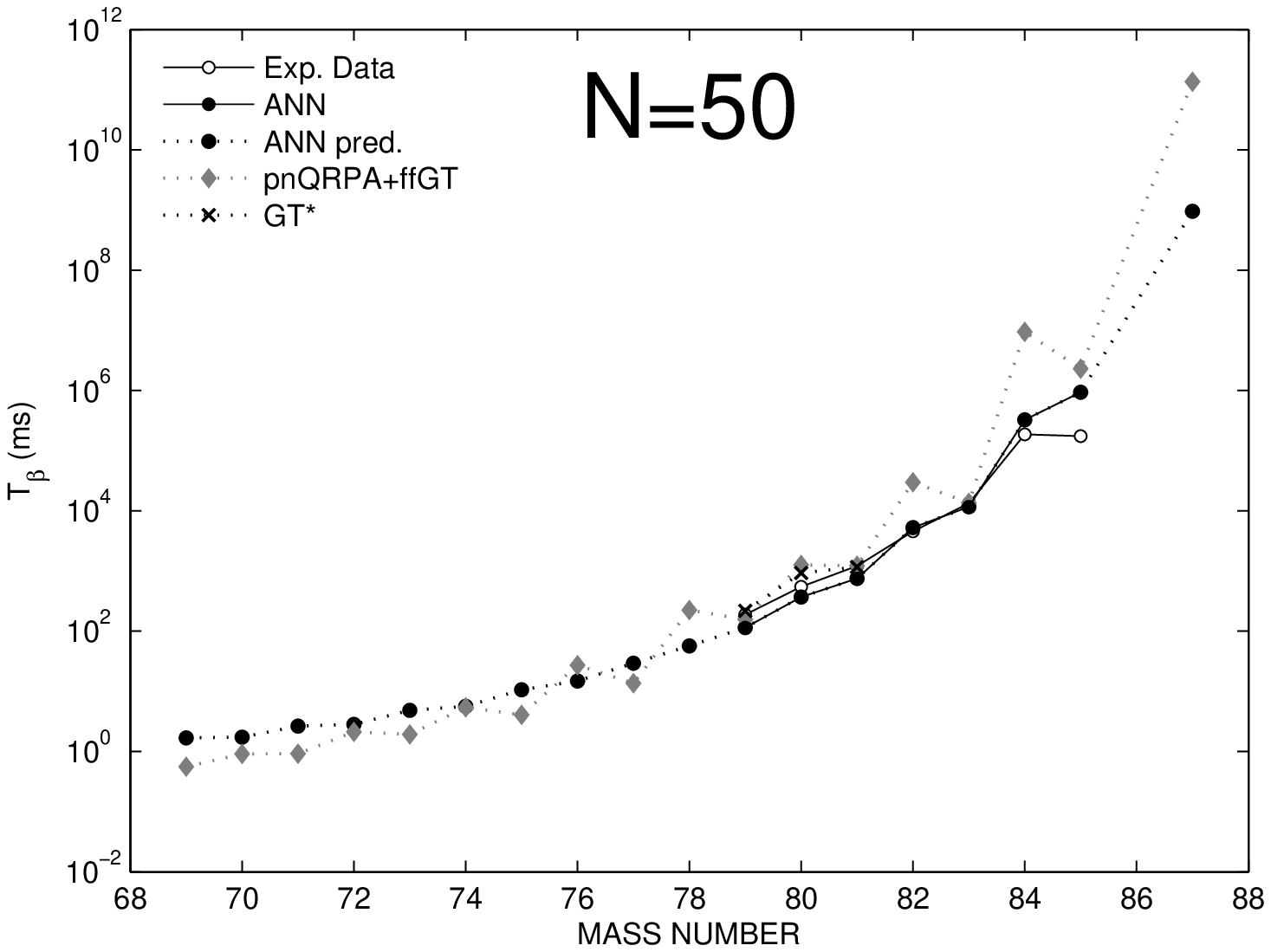}
\caption{\label{fig:n50} The same as in Fig.~\ref{fig:fe26} but for the isotonic chain of $N=50$.}

\noindent{}

\includegraphics[width=3.45in]{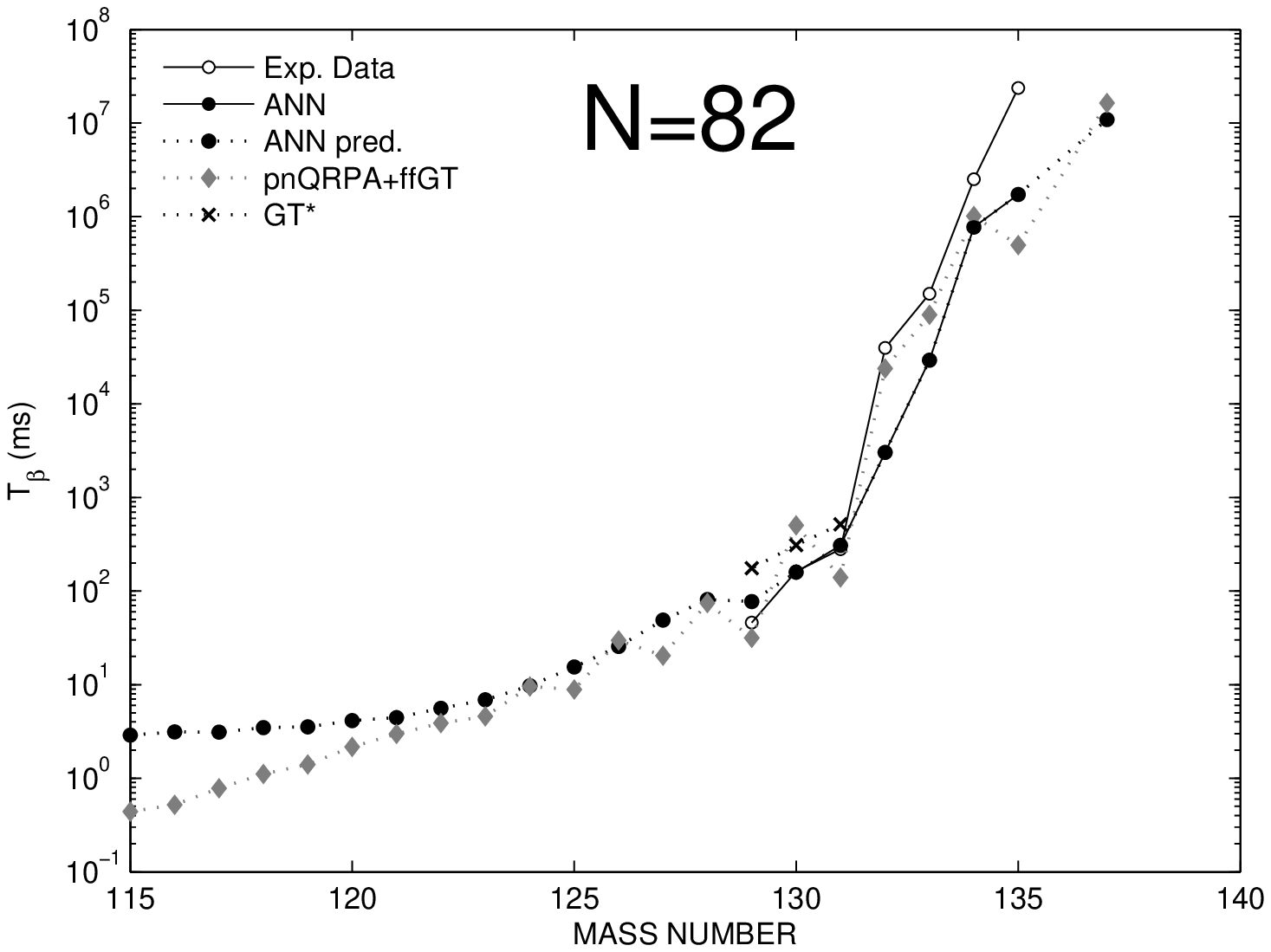}
\caption{\label{fig:n82} The same as in Fig.~\ref{fig:fe26} but for the isotonic chain of $N=82$.}

\noindent{}

\includegraphics[width=3.45in]{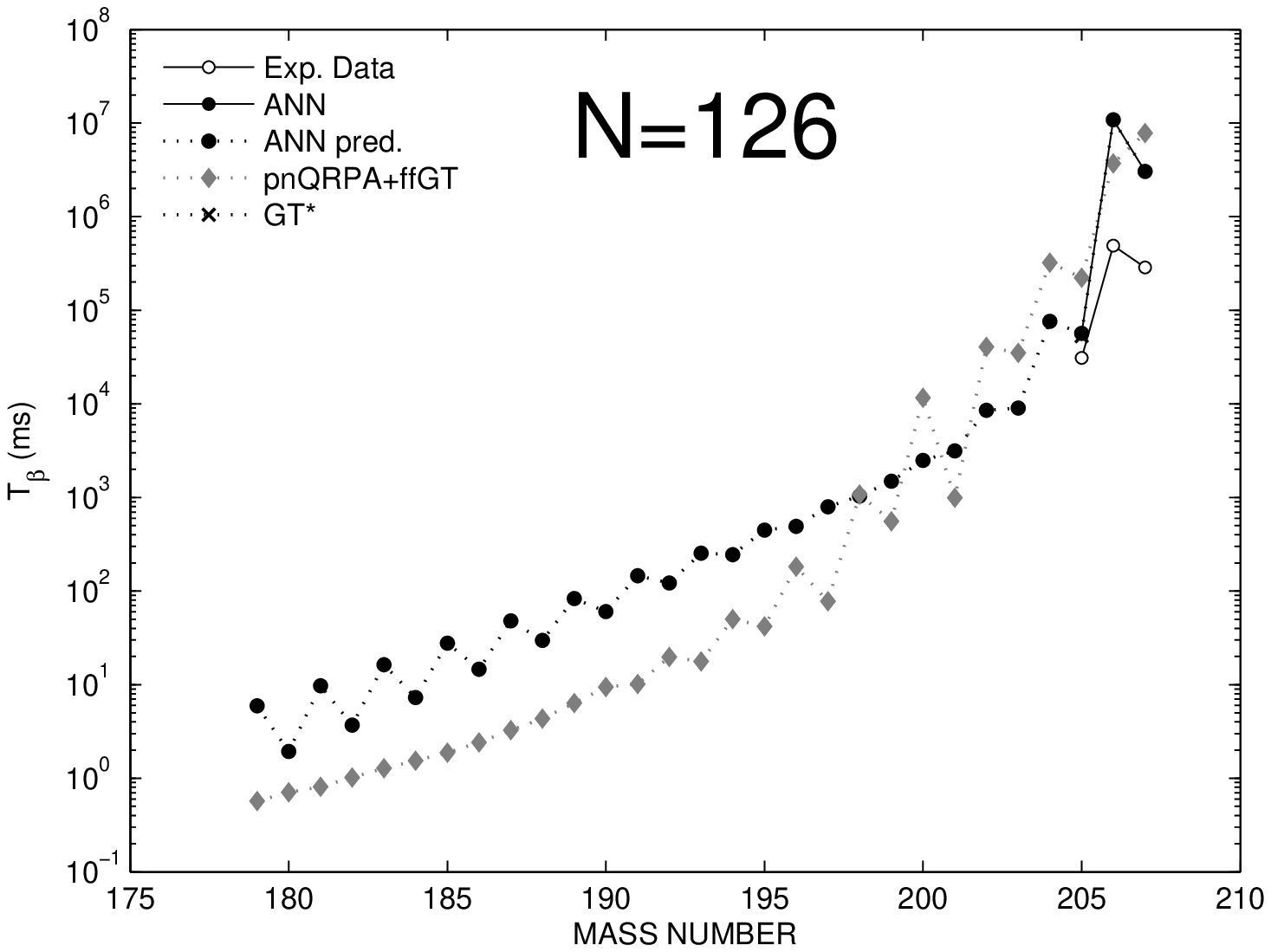}
\caption{\label{fig:n126} The same as in Fig.~\ref{fig:fe26} but for the isotonic chain of $N=126$.}

\end{figure}

\begin{figure}[bht]
\includegraphics[width=3.37in]{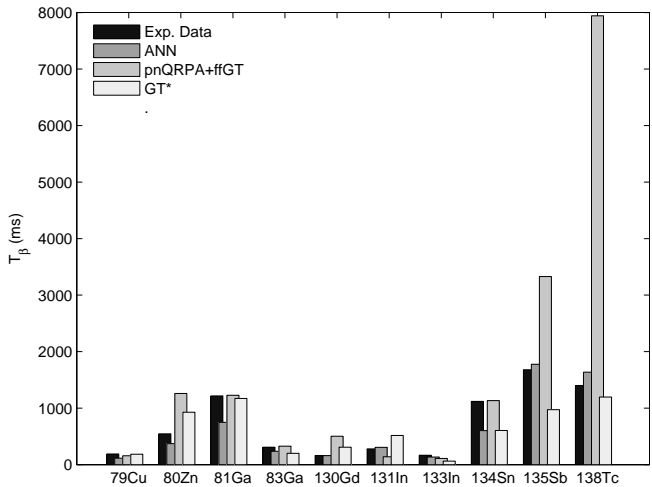}
\caption{\label{fig:fig11} Halflives for $\beta^-$-decaying nuclides that are found near or on a typical r-process path with the neutron separation energy lesser or equal to 3 MeV.}
\end{figure}

Predictions from the ANN model developed here, and improvements upon it,
may prove to be useful for quantitative studies involving r-process
nucleosynthesis. The $\beta$-halflives ($T_\beta$) and
$\beta$-delayed neutron emission probabilities ($P_n$) of those isotopes
lying in the r-process path are the two key $\beta$-decay
parameters that bear upon the $\beta$-strength function
($S_\beta$)~\cite{eirhnh5}.  Accordingly, an approach having global
applicability for accurate prediction of $\beta$ halflives is
needed for detailed dynamical r-process calculations. Moreover, reliable
beta-halflife calculations are of special interest for the r-ladder
isotones $N=50$, $82$, and $126$ where solar abundances peak, since they
determine the r-process time scale.  In
Figs.~\ref{fig:n50}--\ref{fig:n126} we plot the halflives of
closed-neutron-shell nuclei in these significant r-process regions as
predicted by our ANN model, in comparison with corresponding results
from $pn$QRPA+\textit{ff}GT and GT* calculations~\cite{8}.
In particular, it is interesting to compare the various estimates of the
halflife of the doubly magic r-process nucleus $^{78}$\rm{Ni}
($Z=28$, $N=50$).  The result given by the ANN model is consistent
with the recent measurement by Hosmer et al.~\cite{20}.
In Fig.~\ref{fig:fig11}, halflives of $\beta^-$-decaying nuclides that
are found near or on a typical r-process path with neutron separation energy
below 3 MeV are compared with those from $pn$QRPA+\textit{ff}GT and GT*
calculations~\cite{8}.  The results given by the ANN model are
close to the experimental values.

\section{\label{sec:level4}CONCLUSION AND PROSPECTS}

A statistical approach to the global modeling of nuclear properties has been 
proposed and implemented for treatment of the systematics of $\beta^-$
lifetimes of the ground states of nuclei that decay exclusively in this mode.
Specifically, artificial neural networks (ANNs) of multilayer feedforward
architecture are taught to reproduce the experimentally measured
lifetimes of nuclides from a chosen large data set.  Training of the
networks is carried out in such a way that their intrinsic generalization
capabilities can be exploited to predict lifetimes of nuclides outside
the data set used for learning.

We have been able to develop an ANN model of this kind that 
demonstrates very good properties in terms of both the standard
performance measures used in statistical analysis and more
problem-specific quality measures that have been introduced to
assess traditional theoretical models for calculating $\beta^-$
lifetimes on a global scale.  In a purely results-oriented sense
(accurate fitting of given data and good prediction for nuclei not
involved in the fitting process), the performance of this model
matches or surpasses that of traditional models based on 
nuclear theory and phenomenology.  This success opens the
prospect that statistical modeling based on machine learning
can provide a valuable tool in the exploration of $\beta^-$
halflives of newly created nuclei beyond the valley of stability.

Experience gained previously with neural-network modeling
of nuclear systematics (especially the modeling of masses \cite{59,70,29})
strongly suggests that significant further improvements on the current
ANN model of $\beta^-$ systematics are possible, as more sophisticated
training algorithms and machine-learning strategies are continuously
being developed.  Thus we plan further studies along the same lines
with multilayer feedforward perceptrons, while also exploring the potential
of Support Vector Machines.  

It is to be stressed that this program can be no substitute for aggressive
pursuit of traditional, ``theory-thick'' global modeling, which inevitably provides greater insight into the underlying physics responsible for values taken by the targeted nuclear properties. The statistical approach can best
serve in complementary and supportive roles.   We point out that hybrid
statistical-theoretical models show special promise, as demonstrated
in Ref.~\onlinecite{29}.  In that recent work, a 
$\left[{4-6-6-6-1\left|169\right.} \right]$ ANN is
used to model the {\it differences} between measured mass-excess values
and the theoretical values given by the finite-range droplet model (FRDM) of Ref.~\onlinecite{24}, thereby enabling improved prediction of masses away
from stability.

Finally, as this last remark exemplifies, the prospects for fruitful application
of statistical, machine-learning methods extend to a wide range of nuclear 
properties beyond the systematics of $\beta$-decay lifetimes.

\section{\label{sec:level5}ACKNOWLEDGEMENTS}

This research has been supported in part by the U. S. National Science Foundation under Grant No. PHY-0140316 and by the University of Athens under Grant No. 70/4/3309. We wish to thank G. Audi and his team for very helpful communications. JWC is grateful to Complexo
Interdisciplinar of the University of Lisbon and to the Department
of Physics of the Technical University of Lisbon for gracious
hospitality during a sabbatical leave; and to Funda\c{c}\~{a}o para
a Ci\^{e}ncia e a Tecnologia of the Portuguese Minist\'erio da
Ci\^{e}ncia,
Tecnologia e Ensino Superior as well as Funda\c{c}\~{a}o Luso-Americana
for research support during the same period.

\newpage 
\bibliography{paper12bib}

\end{document}